\documentclass[preprint,12pt]{elsarticle}

\usepackage{amssymb}
\usepackage{float}
\usepackage{subcaption}
\usepackage[a4paper, margin=1in]{geometry}
\usepackage{etoolbox}
\usepackage{soul} 
\usepackage{algorithm}
\usepackage{algpseudocode}
\usepackage{amsmath}
\usepackage{amssymb}
\usepackage{array}
\usepackage{float}
\usepackage{color, colortbl}
\definecolor{Gray}{gray}{0.9}
\usepackage{hyperref}
\usepackage{tikz,xcolor,hyperref}
\usepackage{lscape}
\usepackage{longtable}
\usepackage{multirow}
\newcommand{\orcid}[1]{\href{https://orcid.org/#1}{\includegraphics[width=8pt]{orcid.png}}}
\usepackage{amsmath}
\usepackage{booktabs}
\usepackage{caption}
\usepackage{xurl}

\usepackage{pifont}
\usepackage{tabularx}
\usepackage{makecell}

\usepackage[capitalize]{cleveref}

\definecolor{rwGreen}{HTML}{187A32}
\definecolor{rwOrange}{HTML}{D97706}
\definecolor{rwRed}{HTML}{C62828}
\definecolor{rwHighlight}{HTML}{F0F5F8}

\newcommand{\rwYes}{\textcolor{rwGreen}{\ensuremath{\checkmark}}}
\newcommand{\rwPartial}{\textcolor{rwOrange}{\ensuremath{\approx}}}
\newcommand{\rwNo}{\textcolor{rwRed}{\ensuremath{\times}}}

\journal{}

\begin{document}

\begin{frontmatter}



\title{Adaptive LLM-Driven Honeypots vs. LLM-Driven Attackers: Does Narrative Coherence Improve Deception?}


\author[inst1]{Pedro Beltrán-López}

\affiliation[inst1]{organization={Department of Information and Communication Engineering},
            addressline={University of Murcia}, 
            city={Murcia},
            postcode={30100},
            country={Spain}}

\affiliation[inst3]{organization={SCORPION Cybertech},
            addressline={CEEIM, Campus Universitario de Espinardo, 7}, 
            postcode={30100},
            country={Spain}}

\affiliation[inst2]{organization={Department of Applied Mathematics and Computer Science},
            addressline={Technical University of Denmark}, 
            city={Kongens Lyngby},
            country={Denmark}}
            
\author[inst3]{Raul Gonzalez Jimenez}
\author[inst2]{Dario Maddaloni}
\author[inst1]{Manuel Gil Pérez}
\author[inst1]{\corref{cor1}Pantaleone Nespoli}

\cortext[cor1]{Corresponding author}

\begin{abstract}
Large Language Models (LLMs) enable honeypots to generate responses and deceptive artifacts on demand, yet existing approaches evaluate generated content one reply at a time and leave open whether artifacts that jointly describe a coherent event retain an attacker better than unrelated ones. This question becomes pressing as autonomous LLM-driven offensive agents emerge as realistic adversaries. This paper proposes narrative deception, a reactive strategy in which the defender maps each observed attacker command to a MITRE ATT\&CK technique, selects a compatible scenario from a curated catalogue based on MITRE Engage, and deploys mutually referential artifacts (files, processes, services, and logs) on an emulated Ubuntu host while the intrusion is under way. An LLM-powered defender and an autonomous LLM attacker interact through a real SSH session, and the attacker’s termination policy distinguishes abandonment due to suspected deception from abandonment due to loss of interest. A campaign of 1,150 sessions compares four conditions of increasing sophistication (Static, Same-TTP, Multi-TTP, and Chain) across four open-weight attacker models, four defender models, and three attacker skill levels. Engagement increases significantly at every step. With the same artifact budget per reaction, coherent scenarios raise the median session length of advanced attackers from 40 to 165 commands and reduce their abandonment from 86\% to 52\%. The effect grows with attacker skill, holds across all sixteen model pairs, and comes at comparable inference cost. Nearly all sessions ended through loss of interest rather than detection, which suggests that deception should sustain meaningful findings, not only emulator fidelity.

\end{abstract}


\end{frontmatter}



\section{Introduction}
\label{sec:introduction} 
Undoubtedly, cybersecurity can be represented as a fundamental requirement for protecting digital systems, services, and infrastructures against increasingly automated and adaptive threats~\cite{nespoli2026}. In this sense, conventional security mechanisms primarily focus on preventing, detecting, and containing malicious activity. Nevertheless, once an intrusion is underway, defenders may have limited visibility into the adversary's intentions, capabilities, and next actions~\cite{nespoli2018}. Such an inherent characteristic has motivated the development of proactive security techniques that complement conventional defenses by influencing adversary behavior, gathering intelligence, and creating opportunities to observe malicious activity in controlled environments.

Cyber deception represents such approach. Through honeypots, decoys, honeytokens, and misleading artifacts, defenders can create environments that attract adversaries and expose their actions while diverting activity from legitimate assets~\cite{beltran2026}. However, traditional honeypots have relied on predefined command handlers and prepared content, limiting the variety of interactions they can sustain~\cite{moric2025}. In this scenario, Large Language Models (LLMs) offer novel possibilities for overcoming these constraints by generating responses and deceptive artifacts on demand, adapting to observed commands, and maintaining contextual consistency throughout an interaction~\cite{Chaudhry2026, Ersok2025}. Nevertheless, making individual responses plausible does not necessarily make the overall environment convincing or sufficiently engaging. An attacker exploring a host through multiple commands can compare files, processes, services, and logs, potentially identifying inconsistencies or finding little reason to continue the exploitation.

Here, the emergence of LLM-driven offensive agents highlight such a limitation even more. In fact, these agents can plan attack steps, execute commands, interpret their results, and adapt subsequent actions to the information obtained~\cite{LI2026, Shao2026}. Recent studies have begun examining their behavior in deceptive environments, including their command selection, responses to misleading information, and ability to recognize honeypots~\cite{prinos2026honeyquest, maddaloni2026honeyagent}. Nonetheless, existing approaches lack in describing whether artifacts generated deceptively during an intrusion can retain an attacker more effectively when they collectively describe a coherent event rather than appearing as unrelated pieces of information. Furthermore, observing how long an attacker remains connected does not, by itself, explain why a session ends. In fact, an attacker may suspect deception or simply conclude that the target no longer offers useful information. That is, distinguishing these outcomes can be described as crucial for understanding the effectiveness of a defensive, deceptive strategy.

To address this gap, the paper at hand proposes narrative deception, a reactive strategy in which a defender generates and deploys groups of mutually referential artifacts that describe a common fictional event. When an attacker command is associated with a MITRE ATT\&CK technique~\cite{strom2018attack}, the defensive system selects a compatible scenario from a curated catalogue based on MITRE Engage~\cite{mitre_engage} and materializes its artifacts on an emulated Ubuntu host. Specifically, the artifacts share a common subject and are generated together, allowing files, processes, services, and logs to provide mutually reinforcing evidence of the same event. The LLM-powered defender and an autonomous LLM-driven attacker interact through a real SSH session, with the attacker restricted to the information observable through its commands. The presented setting enables the evaluation of narrative deception during an ongoing interaction instead of considering isolated responses or deception prepared exclusively before an attack.

Additionally, an extensive evaluation compares four conditions, i.e., \textsc{Static}, \textsc{Same-TTP}, \textsc{Multi-TTP}, and \textsc{Chain}. Particularly, these conditions progressively introduce reactive deception, artifact diversity, and scenario-based coherence. A campaign of 1,150 completed sessions evaluates these conditions across four attacker models, four defender models, and three attacker skill levels.

Overall, the main contributions of this work can be summarized as follows:
\begin{itemize}
    \item The design of a deceptive system that selects MITRE Engage-based scenarios in response to observed ATT\&CK techniques and deploys coherent, cross-referential artifacts on an emulated Linux host.
    \item The evaluation of LLM-driven attackers and defenders through SSH interactions, comparing four deception conditions and distinguishing session termination due to suspected deception from termination due to loss of interest.
    \item The examination of attacker engagement through 1,150 completed sessions across models and skill levels, including sensitivity analyses of the attacker termination policy and assessments of artifact realism and computational cost.
\end{itemize}

The remainder of this paper is organized as follows. Section~\ref{sec:related_work} reviews related work. Then, section~\ref{sec:design} describes the proposed system, while Section~\ref{sec:setup} presents the experimental setup, and Section~\ref{sec:results} reports and review the results. Next, Section~\ref{sec:discussion} discusses their implications and limitations, and Section~\ref{sec:conclusions} concludes the paper.

\section{State of the Art}
\label{sec:related_work}
This section reviews three related research directions and compares them with our study in Table~\ref{tab:related-work}. The directions cover (1) LLM-based defensive deception~\ref{sec:LLMDef}, (2) LLM-assisted offensive operations~\ref{sec:LLMAtc}, and (3) the behavior of LLM attackers in deceptive environments~\ref{sec:LLMDec}.

\subsection{LLM in defensive deception}
\label{sec:LLMDef}
An SSH honeypot provides an attacker with a decoy host to explore so that the defender can observe the commands issued after the login. In Cowrie \cite{cowrie}, command handlers and a simulated filesystem determine what to respond. The honeypot can keep track of changes during the session, but the commands it supports and its content depend on what was previously prepared. Instead, LLM-based honeypots aim to mitigate this limitation, i.e., a model generates the commands output and file contents on request, and can use earlier interactions to reply consistently with what the attacker has already seen \cite{sladic2024llm}.

ShelLM \cite{sladic2024llm} used an LLM to simulate a Linux shell and asked twelve cybersecurity researchers to identify the parts of its replies that revealed the deception. The researchers knew they were examining a honeypot, so their assessments concerned the plausibility of the answers and the errors that exposed the simulation. HoneyLLM \cite{liuhoneyllm} tested the shell behavior of several language models and deployed the honeypot on the Internet. Commands collected during that deployment were replayed to compare command handling, while connection logs supported a separate comparison of session durations. These durations are difficult to interpret as sustained engagement, since most traffic was automated and the additional delay in generating replies could itself prolong a connection.

AI-enhanced Honeypots \cite{christli2024ai} uses LLaMA-3 to generate SSH responses. The authors compared these responses with Ubuntu outputs using Levenshtein similarity. Commands such as \texttt{pwd} produces close matches, while the responses to \texttt{ps aux} and \texttt{top} differed substantially from the reference output. The authors identify incomplete command coverage and the absence of persistent memory as limitations.

Other studies examine how to maintain consistent responses across commands and sessions. VelLMes \cite{sladic2025vellmes} extends ShelLM into a multi-service framework and retains interaction history so that files and users remain consistent when someone reconnects. The authors tested perceived realism with 89 participants randomly assigned to either a real Ubuntu system or a simulated shell. Participants knew that the host might be a honeypot, and many classified the real Ubuntu system as one, which complicates the interpretation of the results. OHRA \cite{safargalieva2025ohra} also uses session history, but limits its size and combines it with prompts tailored to each protocol. The authors evaluated the generated replies through command replay, similarity measures, and response-time analysis, although the protocol implementations vary in completeness.

ShellBox \cite{yang2025shellbox} uses dependencies between commands to select which parts of the interaction history to include in each reply. When an input tries to make the model abandon its shell role, the system generates a plausible shell error. Its evaluation also uses an LLM to prepare multistage attack sequences that are then executed against real and honeypot systems. Here, the model prepares the attack commands but does not select its next command from the responses received during execution.

Other studies consider how responses or content can be tailored to an attack. HoneyGPT \cite{wang2026honeygpt} combines an explicit record of system state with a shortened interaction history to generate shell replies. The authors report example of prolonged attacker interaction during a three-month field comparison with Cowrie. However, such a  comparison does not isolate whether relationships among generated artifacts contribute to engagement.

LLMs can also generate the content an attacker finds on a decoy host. Beltrán Lopez et al. \cite{lopez2026reactive} redirect suspicious traffic to hosts created on demand and populate them with LLM-generated honeydata suited to the target. The system selects its response according to the security property threatened by the attack and can generate further content as the interaction continues. Its evaluation focuses on operational performance and the similarity of generated documents, without measuring attacker engagement.

These studies show how a honeypot can adapt its replies and content, but each evaluation answers a different question. Output comparisons measure agreement with reference text, while command tests examine how the shell behaves. Participants studies assess whether it appears real, and deployment logs record what attackers did without necessarily explaining why they stayed or left. Bridges et al. \cite{bridges2026sok} also distinguish believability and fidelity from attacker costs, intelligence gained, and defender costs. Our central comparison asks whether generated artifacts lead to greater engagement when they belong to a common scenario than when they are unrelated, with both conditions generating artifacts in response to ongoing attacker activity.

\subsection{LLM as an offensive tool}
\label{sec:LLMAtc}

Early studies examined how LLMs could assist penetration testers and which tasks they could carry out automatically. Happe and Cito \cite{happe2023getting} explored both uses, i.e., they asked a model for testing guidance, and they built a prototype that automatically ran its proposed commands over SSH to attempt privilege escalation on a Linux host. In PentestGPT \cite{deng2024pentestgpt}, the model suggests how to proceed in the attack, and a human operator executes the commands. In the exploratory tests that informed its design, the authors found that models often pursues the most recent finding in depth and overlooked alternative attack paths. That is, PentestGPT records progress in a task tree, helping the model keep track of alternatives as the attack proceeds.

Other systems automate command execution, reducing the need for a human to carry out the model's plan. HackingBuddyGPT \cite{happe2026llms} attempts Linux privilege escalation over SSH. The authors tested twelve cases, each containing a single vulnerability, and compared the results with traditional tools and a professional penetration tester. PenHeal \cite{huang2023penheal} covers vulnerability discovery and remediation recommendations, with an evaluation on Metasploitable2. AutoPentester \cite{ginige2025autopentester} assigns planning, command generation, execution, and result analysis to specialized agents. It was tested on HackTheBox machines and custom vulnerable hosts. Incalmo \cite{singer2026incalmo} extends the setting to emulated multi-host networks, where an LLM plans attacks carried out by specialized agents. Across these studies, the main evaluations use controlled targets with known vulnerabilities and do not examine how the systems respond when an active defender intervenes.

Automating an attack also requires termination mechanisms. PentestGPT's exploratory protocol ends upon a solution or qualitative deadlock, while AutoPentester offers an operator-controlled exit when repetition is detected. PenHeal's evaluation concludes the penetration-testing phase when the tested model detects no additional vulnerabilities. Other mechanisms include reaching root access or an interaction limit in hackingBuddyGPT, and an LLM completion signal or time limit in Incalmo. These rules determine when execution ends, but the evaluations do not distinguish between leaving because the target appears deceptive and leaving because further probing yield nothing new.

The cited evaluations establish capabilities within their respective testbeds, but do not by themselves establish behavior against a defender that changes the available evidence during an intrusion. Our evaluation instead examines engagement within an accessible SSH environment under alternative reactive deception conditions. Continuation and termination are governed by explicit attacker policies, and the resulting measurements must be interpreted within those policies.

\subsection{Attacker LLM against deception}
\label{sec:LLMDec}

Work at the intersection of offensive LLM agents and deception examines both judgments about potential lures and actions taken during attacks. These forms of evidence should be distinguished, i.e., selecting a lead in a questionnaire does not establish that an agent would exploit it, and reporting suspicion does not establish that it would abandon the target.

Honeyquest for LLMs \cite{prinos2026honeyquest} adapts the reconnaissance questionnaire developed by Kahlhofer et al. \cite{kahlhofer2024honeyquest}. The authors compared responses from 21 LLMs with those previously collected from 47 human participants on 174 queries containing genuine vulnerabilities, deceptive leads, and neutral material. Models marked the lines they would pursue or avoid. They selected deceptive leads more frequently than the human participants, and often selected them even in responses that expressed suspicion of a trap. This recognition–action gap indicates that expressed suspicion did not reliably translate into avoidance. However, each query was answered in a single turn, so the study does not show whether an agent would act on those choices or abandon a target during an attack.

Honey for the Agent \cite{maddaloni2026honeyagent} examines how cyber deception influences command selection during SSH sessions. The authors tested nine open-weight models using six prompts covering reconnaissance, deception detection, and exploitation. The experiments comprised 16,200 sessions across a real SSH service and four Cowrie configurations. Model identity was the main determinant of command selection. Deceptive configurations increased the use of targeted commands, namely \texttt{ls} and \texttt{whoami}, even when models were instructed to detect deception. Sessions were capped at 20 or 30 steps and could end earlier under specified termination conditions. The evaluation focused on command patterns and behavioral fingerprints, without analyzing why the attacker abandoned a session.

AdvancedShelLM \cite{sladic2026advancedshellm} uses a Worker to generate shell responses and a Manager to validate them and update an explicit simulated file-system. Its evaluation includes the ARACNE agent interacting with AdvancedShelLM, Cowrie, and genuine Ubuntu under a 20-command limit. ARACNE's classifications were predominantly inconclusive, including on the genuine host. This characteristic limits the conclusions that can be drawn about comparative detectability, while the architecture demonstrates online response generation with maintained state.

HoneyLLMd \cite{fan2026honeyllmd} uses a hierarchical probabilistic automaton to model attack progression and decide which actions to block. An LLM generates deceptive responses that encourage the attacker to explore alternative paths. The authors tested this mechanism with CTFAgent in a controlled challenge involving persistence through a malicious SSH key. The agent followed a prescribed attack process and tried alternative actions when blocked. The evaluation therefore examines how adaptive responses influence attack behavior, without comparing coherent and unrelated artifacts or analyzing why the attacker abandoned the attack.

CHeaT \cite{ayzenshteyn2025cloak} evaluates deception specifically designed for LLM attack agents. Its Cloak, Honey, and Trap strategies misdirect agents, expose them through lures and honeytokens, and delay or disrupt their progress. GPT-4o assisted payload preparation; experiments include several agent frameworks and model backend, with a separate end-to-end evaluation using PentestGPT. Some techniques already use linked breadcrumbs or cyclic references across deceptive artifacts. The evaluated payloads are prepared before the attack; strategic introduction of additional traps during an ongoing attack is identified as future work.

Taken together, the three lines of work leave the same question open from different sides. LLM honeypots make each reply more realistic and keep the host consistent across commands and sessions, but they evaluate replies one at a time, through similarity to reference outputs, human judgments, or deployment logs, and none of them relates the content it generates into a single event. The systems that adapt their deception to the attack, such as reactive decoy population \cite{lopez2026reactive} and HoneyLLMd \cite{fan2026honeyllmd}, choose what to present according to the attack, but their artifacts do not describe a common event, and their evaluations do not compare coherent with unrelated deception. LLM attack agents are evaluated on controlled targets by whether they complete a task, so their stopping rules never depend on what a defender offers. Finally, the studies of LLM attackers facing deception use lures prepared before the attack or isolated cues, with a single kind of attacker, and they record what the attacker did without separating whether it left because it suspected a trap or because it found nothing new.

The contribution of this work is narrative deception: the defensive system answers each technique it identifies with a scenario, a short event told by several artifacts that share one subject and corroborate one another, chosen at run time from a public defensive framework, MITRE Engage, and written while the attack is under way. Prior work contains some of the ingredients, as Table~\ref{tab:related-work} shows. Adaptive deception exists, linked breadcrumbs exist in prepared form \cite{ayzenshteyn2025cloak}, LLM attackers have been placed in front of deception, and some agents stop when their progress stalls. What no prior study does is build the scenario while the attack is under way and test whether it matters, that is, compare in the same environment and against the same attackers artifacts that tell one event with artifacts that are unrelated. Our evaluation is designed for that comparison, with a ladder of four conditions in which the last two plant the same budget of artifacts per reaction, attackers of three skill levels that value different evidence, and a separate record of the sessions ended by detection and by loss of interest. Table~\ref{tab:related-work} groups the reviewed studies by what they propose, namely deceptive systems, attack agents, and studies of attackers facing deception, and compares them on the mechanism of deception and on how it is evaluated.

\begin{table*}[h!]
\centering
\caption{Reviewed studies grouped by what they propose and compared on the mechanism of deception and on how it is evaluated. \textit{Adaptive:} attacker actions guide which lures, decoys, or misleading paths are presented during the session; shell replies and state updates alone are excluded. \textit{Narrative:} the artifacts of one reaction describe a common event and refer to one another. \textit{LLM attacker:} an autonomous LLM attacker interacts with the deception. \textit{Leaving criteria:} termination linked to suspected deception or stalled progress, beyond fixed budgets or task completion alone. \textit{Coherence comparison:} coherent and unrelated reactive artifacts are compared in the same environment. The first four columns are shared with prior work, and this work reuses them as the setting in which narrative deception is tested.}
\label{tab:related-work}
\begingroup
\footnotesize
\setlength{\tabcolsep}{2pt}
\renewcommand{\arraystretch}{1.12}

\begin{tabularx}{\textwidth}{
    @{}
    >{\raggedright\arraybackslash}p{0.15\textwidth}
    c
    >{\raggedright\arraybackslash}X
    *{5}{>{\centering\arraybackslash}p{0.09\textwidth}}
    @{}
}
\toprule
\textbf{Work} &
\textbf{Ref.} &
\textbf{Study focus} &
\makecell[t]{\textbf{Adap-}\\\textbf{tive}} &
\makecell[t]{\textbf{Narra-}\\\textbf{tive}} &
\makecell[t]{\textbf{LLM}\\\textbf{attacker}} &
\makecell[t]{\textbf{Leaving}\\\textbf{criteria}} &
\makecell[t]{\textbf{Coher.}\\\textbf{comp.}} \\
\midrule

\multicolumn{8}{@{}l}{\emph{Deceptive systems}} \\

ShelLM &
\cite{sladic2024llm} &
Linux-shell emulation &
\rwNo & \rwNo & \rwNo & \rwNo & \rwNo \\

HoneyLLM &
\cite{liuhoneyllm} &
Contextual shell emulation &
\rwNo & \rwNo & \rwNo & \rwNo & \rwNo \\

AI-enhanced honeypots &
\cite{christli2024ai} &
Shell-output fidelity &
\rwNo & \rwNo & \rwNo & \rwNo & \rwNo \\

VelLMes &
\cite{sladic2025vellmes} &
Multi-service emulation &
\rwNo & \rwNo & \rwNo & \rwNo & \rwNo \\

OHRA &
\cite{safargalieva2025ohra} &
Multi-protocol emulation &
\rwNo & \rwNo & \rwNo & \rwNo & \rwNo \\

ShellBox &
\cite{yang2025shellbox} &
Shell consistency and robustness &
\rwNo & \rwNo & \rwPartial & \rwNo & \rwNo \\

HoneyGPT &
\cite{wang2026honeygpt} &
State-aware shell emulation &
\rwNo & \rwNo & \rwNo & \rwNo & \rwNo \\

Reactive deception &
\cite{lopez2026reactive} &
Redirection and decoy population &
\rwYes & \rwNo & \rwNo & \rwNo & \rwNo \\

AdvancedShelLM &
\cite{sladic2026advancedshellm} &
Stateful SSH deception &
\rwNo & \rwNo & \rwYes & \rwNo & \rwNo \\

HoneyLLMd &
\cite{fan2026honeyllmd} &
Attack-path steering &
\rwYes & \rwNo & \rwYes & \rwNo & \rwNo \\

CHeaT &
\cite{ayzenshteyn2025cloak} &
Deception against LLM agents &
\rwNo & \rwPartial & \rwYes & \rwPartial & \rwNo \\

\midrule

\multicolumn{8}{@{}l}{\emph{LLM attack agents}} \\

PentestGPT &
\cite{deng2024pentestgpt} &
Penetration-testing assistance &
\rwNo & \rwNo & \rwNo & \rwPartial & \rwNo \\

PenHeal &
\cite{huang2023penheal} &
Pentesting and remediation &
\rwNo & \rwNo & \rwNo & \rwYes & \rwNo \\

hackingBuddyGPT &
\cite{happe2026llms} &
Linux privilege escalation &
\rwNo & \rwNo & \rwNo & \rwNo & \rwNo \\

AutoPentester &
\cite{ginige2025autopentester} &
Automated penetration testing &
\rwNo & \rwNo & \rwNo & \rwPartial & \rwNo \\

Incalmo &
\cite{singer2026incalmo} &
Multi-host red teaming &
\rwNo & \rwNo & \rwNo & \rwNo & \rwNo \\

\midrule

\multicolumn{8}{@{}l}{\emph{LLM attackers facing deception}} \\

Honeyquest for LLMs &
\cite{prinos2026honeyquest} &
Reconnaissance judgments &
\rwNo & \rwNo & \rwPartial & \rwNo & \rwNo \\

Honey for the Agent &
\cite{maddaloni2026honeyagent} &
Command selection under deception &
\rwNo & \rwNo & \rwYes & \rwPartial & \rwNo \\

\midrule

\textbf{This work} &
&
\textbf{Narrative deception and its effect on engagement} &
\rwYes & \rwYes & \rwYes & \rwYes & \rwYes \\

\bottomrule
\end{tabularx}

\par\smallskip
\begin{minipage}{\textwidth}
\scriptsize
\raggedright
\textit{Legend:}
\rwYes\ Addressed
\rwPartial\ documented but restricted treatment
\rwNo\ not integrated.
\end{minipage}

\endgroup
\end{table*}

\section{System Design}
\label{sec:design}

This section describes the architecture of the proposed system, in which a language-model defender and a language-model attacker confront each other through a real SSH session over an emulated host. Section~\ref{subsec:architecture} presents the overall structure and the isolation constraints that govern the interaction between both agents. Section~\ref{subsec:defender} then details the defensive pipeline, from deciding what the host claims to be to creating deceptive artifacts inside it, whereas Section~\ref{subsec:attacker} describes the attacker, covering how it plans its actions and how it decides whether a target remains worth its time. The implementation of both systems is publicly available~\cite{llmhoney_engage}.

\subsection{Architecture overview}
\label{subsec:architecture}

\begin{figure}[H]
    \centering
    \includegraphics[width=\textwidth]{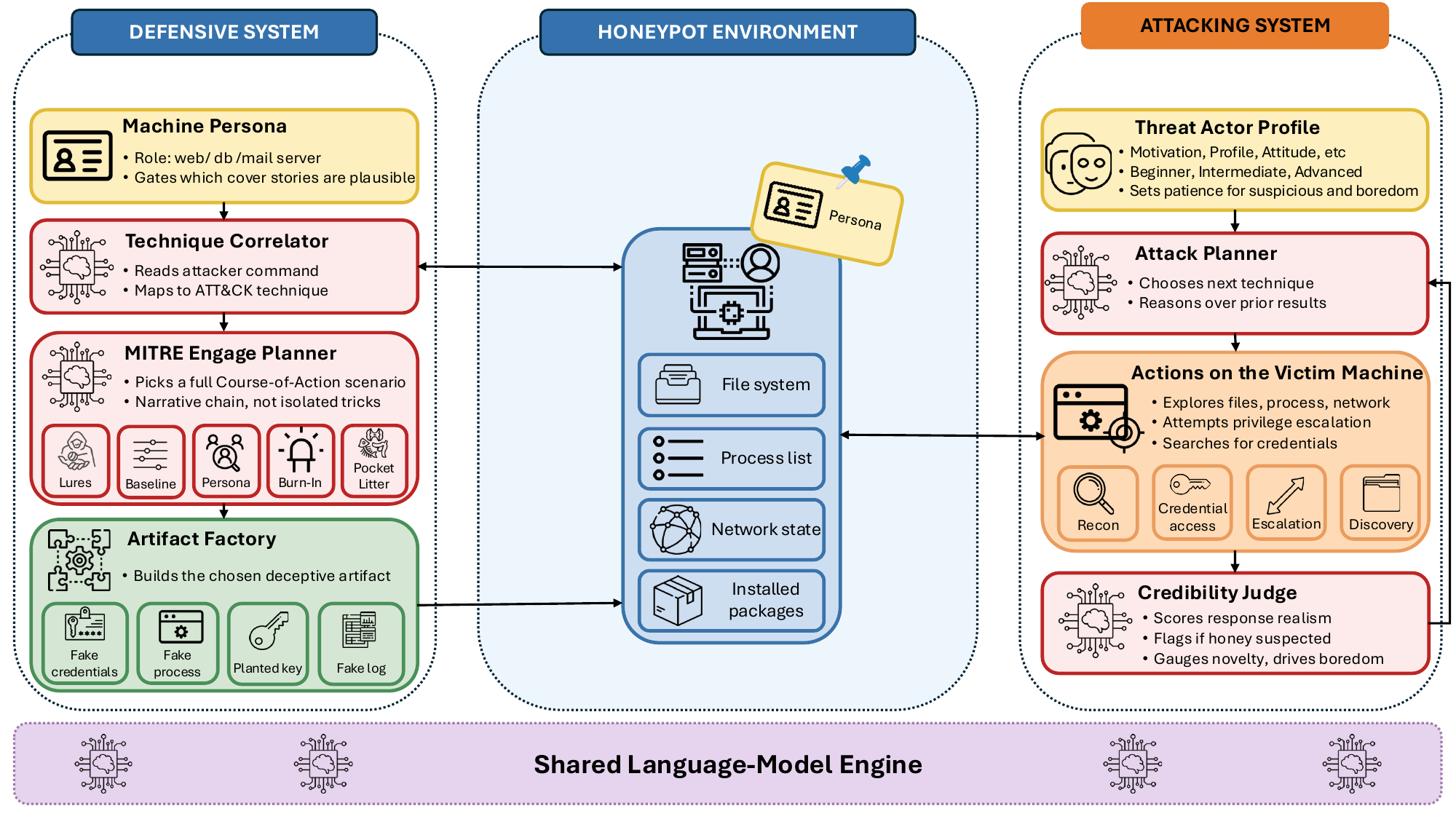}
    \caption{Architecture of the proposed system. The attack phases and Engage activities shown are representative examples; Section~\ref{subsubsec:actions} describes all the tactics used.}
    \label{fig:architecture}
\end{figure}

The system comprises three parts, shown in \figurename~\ref{fig:architecture}, i.e., a defensive system on the left, an attacking system on the right, and between them the honeypot environment, which is the only thing the two have in common. Their interaction follows a closed loop. Specifically, the attacker, guided by a profile that fixes who it is and what it wants, decides on its next move and issues the corresponding command to the host. The environment answers that command as a real server would. In parallel, the defensive system observes the same command, interprets what the attacker is trying to achieve, chooses a deception response consistent with what the host claims to be, and turns that response into new artifacts inside the environment, so that what the attacker finds next already reflects the defense's reaction. Rather than planting isolated tricks, that response takes the form of a \emph{scenario}: a short, plausible event told by several artifacts that all refer to the same fictional event, selected from a curated catalogue. The attacker, in turn, examines each answer it receives, judging both whether it looks genuine and whether it offers anything worth pursuing, and uses that judgment to plan its following step or to leave. The remainder of this section describes each component of that loop, and Sections~\ref{subsec:defender} and~\ref{subsec:attacker} then detail how each one has been implemented.

The two agents interact exclusively through a real SSH session, the channel through which an attacker usually reaches a Linux server, and which gives the attacker exactly what a real one would get, namely the textual output of its commands. Letting both agents share an API or a simulator would be simpler, but it would expose information no real attacker has and would make the conclusions impossible to transfer to real attacks. The separation holds in both directions: the attacking system never learns which capability was deployed, or whether a reaction occurred at all, and the defensive system never sees the attacker's profile, plan, skill level, or judgments, only the commands that arrive through the session, exactly as a deployed honeypot would. Any behavioral difference between conditions must therefore originate in what the target showed the attacker, never in privileged knowledge of how it was configured, and since both agents run as separate operating system processes connected over SSH on \texttt{127.0.0.1:2222}, this isolation is enforced by process boundaries rather than by convention.

The attacking system is not designed as a honeypot detector. It models an attacker that spends effort on a target only while the target keeps rewarding it, as a real attacker with finite time does, so its judgment has two sides: whether the responses look genuine, and whether they offer something worth pursuing, such as credentials, services, or data. A target can fail on either side, and both failures matter to the defense, since in both cases the attacker leaves. Measuring how long the attacker stays and why it stops is therefore what allows deception strategies to be compared.

What kind of host the defensive environment claims to be is governed by its \emph{machine persona}, shown above the environment in \figurename~\ref{fig:architecture}: the role the server plays, with the packages and services that role implies, fixed for the whole session so that everything the attacker observes belongs to the same machine. The persona describes the host, not the attacker, and should not be confused with the threat actor profile of Section~\ref{subsubsec:profile}; Section~\ref{subsubsec:persona} details how it is selected and how it constrains what the defender plants.

Although the evaluation pairs one attacker with one defender, every SSH connection receives its own session, emulated host, machine persona, and deception state, so a single defensive system can serve several attackers at once without any of them seeing what was planted for the others. Sessions in this work nevertheless run one at a time, for the reason given in Section~\ref{subsec:campaign}, and attackers that coordinate against one target, or defenders that share one scenario across hosts, are left as future work.

\subsubsection{Honeypot environment}
\label{subsubsec:environment}

The honeypot environment deserves its own description, since its fidelity bounds what the whole system can demonstrate. In the proposed architecture, it is an emulated Ubuntu host whose file system, process list, network state, and installed packages constitute everything the attacker can observe. Rather than wrapping a shell around a real machine, the environment implements a virtual file system, a process table, network interfaces and connections, systemd units, and a package database, all of which respond to standard commands with plausible output. Ninety-one commands are implemented natively, covering file inspection, process and service management, networking, package management, and privilege changes, and the environment additionally incorporates the command implementations of the Cowrie honeypot~\cite{cowrie}. A command outside that set is answered as a real shell answers a missing binary, namely with \texttt{command not found}. It is also important to state that whether such coverage suffices was measured rather than assumed. Across the evaluation campaign, fewer than $2\%$ of the commands the attacker issued exposed a genuine gap in the emulation, chiefly multi-line shell constructs and a handful of networking binaries, while about a further $4\%$ requested software that a stock Ubuntu server would not have installed either, such as PowerShell cmdlets or cloud provider tools, and therefore received exactly the answer a real host would give. The output format of the most common commands was also compared with that of a genuine Ubuntu~22.04 host, and Section~\ref{subsec:realism} reports where the emulator still departs from it. In this sense, emulation, rather than a real sandboxed host, is what permits the defender to fabricate arbitrary state on demand, i.e., a genuine machine cannot produce a fake log file with a plausible history, nor inject a process that was never executed.

Besides, the environment keeps state, and this characteristic matters to both agents. On the defensive side, every artifact the defender plants must survive contact with an actual shell, so a fabricated log file has to appear under \texttt{ls}, be readable with \texttt{cat}, and carry plausible timestamps and ownership, exactly like any other file. On the attacking side, what the attacker does changes the host as well, since a file it creates or an account it adds genuinely exists afterwards and shows up in later commands. Every session starts from a fresh copy of the host, so nothing planted or created in one session is visible in the next.

\subsubsection{Shared language-model engine}
\label{subsubsec:engine}

Both agents delegate their reasoning to language models, and \figurename~\ref{fig:architecture} places those model calls on a shared engine at the bottom of the diagram. Five components draw on that engine: the Technique Correlator, the MITRE Engage Planner, and the Artifact Factory on the defensive side, and the Attack Planner and the Credibility Judge on the attacking side, where the Attack Planner makes two kinds of call, one to choose the next technique and one to choose follow-up commands. Since sharing anything between attacker and defender might suggest a channel through which one could influence the other, it is worth stating precisely what is shared and what is not.

What both sides share is infrastructure, namely a single local inference server and a common pool of open-weight models of comparable size (Section~\ref{subsec:models}), and no information. Every model call is an independent, stateless request that carries only the prompt of the role issuing it, the defensive roles build their prompts only from the commands observed in the session, and the attacking roles only from the outputs they receive, so nothing produced by one side reaches the other except through the SSH channel. The pool is shared for experimental control: every model plays both roles across the campaign, so neither side holds a systematic reasoning advantage. If the shared engine allowed one agent to influence the other, this influence would be strongest when both agents run the same model, and the effect of the deception conditions would then differ between same-model and different-model pairs. The results show no such difference: in each of the sixteen ordered model pairs, including the twelve in which the defensive system and the attacker run different models, abandonment never increases from one deception condition to the next, as Section~\ref{subsec:generalisation} reports.

\subsection{Defensive System}
\label{subsec:defender}

The defensive system practices narrative deception: rather than serving static content, it answers each technique it identifies with a scenario, a set of artifacts that describe the same fictional event and corroborate one another, so that what the attacker finds next responds to what it is doing and agrees with what it has already seen. Each command triggers a pipeline of four stages, arranged vertically in the left column of \figurename~\ref{fig:architecture}, running from interpreting what the attacker just executed to creating concrete artifacts inside the emulated host. Timing deserves particular attention at this point, because an ideal deceptive host would answer every command as quickly as a genuine one, and any added delay is a possible clue that the host is not genuine. Executing a command and producing its response therefore never involve a language model in our proposal, and the emulated host computes the output in under a millisecond, at the same cost in every experimental condition. Generating the deceptive artifacts, on the contrary, takes seconds, since it requires model calls. This generation therefore runs in the background: the host answers each command at once, the attacker keeps working while the defensive system prepares its reaction, and the artifacts appear in the host as soon as they are ready. Whether the attacker sees them depends on what it does next, but it usually does, for two reasons. The defensive system reacts to the part of the host the attacker is working on, and a scenario plants its artifacts in several places at once, such as a log, a process, a service, and a file, so finding it does not depend on any single later command. This is one more advantage of scenarios over isolated artifacts: an isolated artifact that appears after the attacker has looked in its only location is lost, whereas the other artifacts of a scenario remain to be found. Section~\ref{subsec:cost} quantifies these costs in detail.

The central unit of every reaction is the \emph{scenario template}. Imagine a defensive environment that answers each command with an isolated artifact, a series of unrelated fake logs or fake processes. Such an environment can produce a great deal of deception that, nonetheless, describes no common event, and an attentive attacker notices that the pieces do not belong together. A scenario template instead describes a short, plausible event, such as a backup job that failed and left a log and a temporary file behind, together with the ordered chain of capabilities that creates it, so that every artifact planted in a single reaction refers to the same fictional event and corroborates the others. The catalogue contains fifty-seven such templates, written for this work, which cover twenty-two MITRE Engage activities with chains of $3.44$ capabilities on average. The stages that follow decide which template applies and how it is realized, i.e., the Machine Persona restricts the templates that are plausible on the host, the Technique Correlator identifies what the attacker is attempting, the MITRE Engage Planner selects the template that responds to it, and the Artifact Factory turns that template into concrete artifacts sharing a common subject. Section~\ref{subsubsec:engage} defines scenario templates formally.

Four terms are used in a precise sense in the rest of the paper. A \emph{scenario} is one such fictional event together with the artifacts that tell it, and \emph{narrative deception} is the approach of answering every identified technique with a scenario rather than with isolated artifacts. \emph{Coherence} is the property the defensive system builds into each scenario, and it works at three levels: at the level of names, every artifact of a scenario carries one shared \emph{subject}; at the level of structure, the capabilities are applied in a fixed order and only on hosts whose role makes them plausible; and at the level of narrative, the artifacts are written together from a description of the event, so that their free text can refer to one another. \emph{Corroboration}, finally, is the part of that coherence the attacker can check for itself, namely discoveries that share their name with other discoveries, and it is the only level that the rules of the attacker measure (Section~\ref{subsubsec:judge}).

\subsubsection{Machine Persona}
\label{subsubsec:persona}

Every session begins by drawing a machine persona, which fixes what the host claims to be for the duration of that engagement. A persona specifies a role, such as web server, database server, or mail server, together with the packages, listening services, and kinds of scenario that the role makes plausible. Formally, a persona $p$ is a tuple:

\begin{equation}
    p = \langle r,\; P_r,\; S_r,\; T_r \rangle
    \label{eq:persona}
\end{equation}

\noindent where $r$ denotes the role, $P_r$ the set of installed packages, $S_r$ the services and ports exposed, and $T_r$ the subset of deceptive scenario templates (introduced in Section~\ref{subsubsec:engage}) that are plausible on a host playing that role. Within the scope of the proposed work, six roles are defined: a web server, a dedicated database server, a CI/CD build runner, an internal file and backup server, an outbound mail relay, and a developer workstation. These roles cover the main functions a Linux server performs in an organization, and each holds a different kind of asset that an attacker looks for once inside: web application secrets, database contents, build and deployment credentials, backups, mail, and developer keys. Adding a role requires no change to the pipeline, since a role is defined only by its packages, its listening ports, and the scenario tags it admits, and because fifty-one of the fifty-seven templates are universal, a new role starts with almost the whole catalogue available. The selection is deterministic given the session identifier $\text{sid}$, a unique label assigned to every SSH connection, and a global seed, derived through a SHA-256 digest~\cite{fips180-4} so that any session can be reproduced exactly while different sessions receive uncorrelated personas.

To illustrate with an example, a session that draws the database server persona has $r$ set to a dedicated database server, $P_r$ containing \texttt{postgresql-14}, \texttt{pgbouncer}, and \texttt{redis-server} at realistic Ubuntu package versions, $S_r$ exposing PostgreSQL on port $5432$ and Redis on port $6379$, and $T_r$ admitting every generic scenario together with those specific to a database host, such as a decoy backup script and the dump it supposedly produced, while excluding those that only make sense elsewhere, such as a mail transport agent or a developer's personal dotfiles. From the first command onwards, the host is consistent with that role across every view an attacker may take of it, e.g., \texttt{dpkg -l} lists the database packages, \texttt{ss -tlnp} shows both ports listening, and \texttt{systemctl} reports the corresponding services as running. Any deception planted later must fit the same picture, so such a host may reveal a stale database dump, but never a mail queue, for example.


Beyond populating the initial environment, the persona acts as a filter over everything the defensive system plants later. A scenario that would be implausible on the drawn role, such as database credentials on a host running no database or a mail queue on a machine with no mail transport agent, would betray the deception more readily than planting nothing at all. Therefore, each scenario template declares the roles it suits, or none when it is plausible on any server, and before making any selection the MITRE Engage Planner in Section~\ref{subsubsec:engage} discards every candidate whose declared roles exclude the active persona. If no candidate survives, the defensive system deploys a single capability instead of a scenario. This can happen only for the few activities that have no template for the active role, and it did not occur in the campaign, where every \textsc{Chain} reaction was resolved with a scenario template. Since most templates fit any role, the filter removes few options and prevents only the inconsistencies that would give the host away.

\subsubsection{Technique Correlator}
\label{subsubsec:correlator}

Once a command arrives, the correlator determines what the attacker is attempting in adversarial terms. A language model receives the command text, its recent history, and the current environment state, then returns a MITRE ATT\&CK technique identifier~\cite{strom2018attack} together with a confidence in that assignment, whose use is described below. Writing $c_j$ for the $j$-th command of the session and $H_j$ for the history of preceding commands, the correlator implements:

\begin{equation}
    \phi: (c_j,\; H_j) \mapsto (\tau_j,\; \omega_j), \qquad \tau_j \in \mathcal{T},\; \omega_j \in [0,1]
    \label{eq:correlator}
\end{equation}

\noindent where $\phi$ denotes the correlation function realized by the language model, $\tau_j$ the ATT\&CK technique it assigns to the command, $\mathcal{T}$ the set of ATT\&CK techniques known to the system, and $\omega_j$ the confidence of that assignment. For instance, the command \texttt{echo "* * * * * /tmp/evil.sh" > /tmp/persistevil \&\& crontab /tmp/persistevil} writes a schedule line to a file and installs it with \texttt{crontab}, and the correlator maps it to T1053.003, \emph{Scheduled Task/Job: Cron}, a persistence technique, regardless of the particular file names involved.


Command history matters more than it might first appear, which is why the correlator receives the five most recent commands together with the one it classifies. For example, a single \texttt{cat auth.log} is ambiguous in isolation: issued after \texttt{cd /var/log} and \texttt{ls}, it belongs to a survey of the file system, whereas the same read following a failed \texttt{sudo} attempt and a search for SSH keys belongs to a hunt for credentials. Because the correlator sees this context, the defensive system answers the goal the attacker is pursuing rather than the command in isolation, planting credentials when the attacker is searching for them and not every time it reads a file. The approach has three limitations. A wrong classification carries over to the following steps, so the system would answer the wrong question convincingly. The window is deliberately short, which keeps prompts and latency bounded but means that intent spread over a longer stretch of the session is only partly visible. Finally, the first commands of a session have little or no history to draw on, and are therefore classified on their own. Section~\ref{sec:discussion} outlines how each of them could be addressed.

Every command the attacker issues, apart from a handful of generic reconnaissance commands described below, goes through the correlator, which always attempts to associate it with an ATT\&CK technique. The confidence $\omega_j$ expresses how clearly the command points to the technique, and whenever $\omega_j \geq \lambda$, with $\lambda = 0.6$, the command is associated with the technique and the defensive system reacts. Commands that are not malicious as such, as with routine navigation or a typing error, or that are unfamiliar, as a previously unseen exploit would be, typically fall below that threshold. They are then simply executed and answered as a genuine host would answer them, and no deception is planted, which is also what happens when the technique has no counterpart in the MITRE Engage catalogue. Consequently, ambiguity costs the defender an opportunity to react, but it never produces a response that does not fit the host, a limitation to which Section~\ref{sec:discussion} returns. A previously unknown exploit does not escape the mechanism entirely either, because ATT\&CK describes behavior rather than vulnerabilities: whatever the attacker does once the exploit succeeds, such as establishing persistence or harvesting credentials, remains recognizable, so the defender regains the opportunity to react at that point.

Two engineering decisions constrain this stage, both motivated by model capacity rather than classification quality. First, nine generic reconnaissance commands, such as \texttt{ps aux}, \texttt{who}, or \texttt{ls -la /tmp/}, bypass correlation, since they are issued constantly against any server, always map to the same discovery techniques, and would otherwise flood the analysis queue. Such an exception applies identically in every experimental condition and can only reduce the defender's opportunities to react. Second, the number of correlation calls that can run at the same time is limited, because running too many at once slowed down the part of the system that serves the SSH connection and could drop it, which no real host would do.


\subsubsection{MITRE Engage Planner}
\label{subsubsec:engage}

Knowing which technique the attacker is using does not tell the defender how to respond. Such a decision is made by the MITRE Engage Planner, the component that holds one of the central contributions of this work. Its output is a scenario, i.e., a short, believable event that the host tells the attacker through several artifacts at once, such as a file, a log entry, and a running process that all refer to the same occurrence. The planner builds that scenario in four steps, described in turn below:
\begin{enumerate}
    \item it finds the defensive activities that answer the identified technique;
    \item it turns those activities into concrete operations on the host, called capabilities;
    \item it organises those operations into scenario templates;
    \item it picks the template that best suits this attacker.
\end{enumerate}
Two further mechanisms, described after the fourth step, complete the planner: a shared subject keeps the artifacts of a scenario consistent with one another, and a record of the scenarios already deployed keeps the host from repeating itself.

The first step goes from the technique to the defensive activities that can answer it. The planner looks up which defensive activities can answer the identified technique $\tau \in \mathcal{T}$. It uses the mapping published by MITRE Engage~\cite{mitre_engage}. Writing $\mathcal{A}$ for the set of Engage activities, this step returns the subset $A(\tau) \subseteq \mathcal{A}$ that MITRE links to $\tau$. Typical activities are \emph{Lures}, which plants decoys meant to attract the attacker, \emph{Pocket Litter}, which scatters the everyday clutter a real machine accumulates, and \emph{Software Manipulation}, which changes what system tools report. For example, a technique aimed at stealing credentials maps naturally onto \emph{Lures}: an attacker looking for secrets should find some. Using the official mapping, rather than one designed for this work, means that every reaction rests on a public, documented link between what attackers do and how defenders can respond.

The second step turns those activities into concrete operations, since MITRE Engage activities are abstract. That is, \emph{Lures} says that decoys should be planted, but not which file to create or what to write in it. To make them executable, twenty-two of the twenty-three activities were linked to a set of \emph{capabilities}; the remaining one, \emph{Hardware Manipulation}, has no counterpart on an emulated host. A capability $k \in \mathcal{K}$ is a single, configurable operation that the defensive system can perform on the emulated host, such as creating a log file, adding a process to the process list, or planting an SSH key. The link works in both directions: one activity can offer several capabilities, and one capability can serve several activities. The number varies widely, from the fourteen capabilities of \emph{Lures}, which range from fake cloud credentials to planted SSH keys, to the single capability of eleven of the activities, an imbalance whose consequences are discussed in Section~\ref{sec:setup}. This layer is what connects an abstract defensive framework to real operations on a host, and it is what allows MITRE Engage to drive a running system rather than serve only as a planning vocabulary.

The third step organizes those operations into scenarios. Running capabilities one at a time, or picking a few at random, would plant artifacts that each look plausible but have nothing to do with one another. That is exactly the deception used as a control in the experiments (Section~\ref{subsec:conditions}). The proposed planner instead works with complete scenarios. A scenario template $\sigma$ is defined as:

\begin{equation}
    \sigma = \big\langle a,\; \langle k_1, k_2, \ldots, k_n \rangle,\; \nu,\; R_\sigma \big\rangle
    \label{eq:scenario}
\end{equation}

\noindent where $a \in \mathcal{A}$ is the MITRE Engage activity the scenario answers, $\langle k_1, \ldots, k_n \rangle$ with $k_i \in \mathcal{K}$ is the ordered list of capabilities that builds it, $\nu$ is a one-sentence description, in natural language, of the event the resulting artifacts tell together, and $R_\sigma$ is the set of machine roles on which the scenario makes sense, left empty when it fits any role. The narrative $\nu$ is not executed: it is given to the language model when the scenario is chosen and when its artifacts are generated, and it also determines the subject of the scenario (Equation~\ref{eq:subject}). The list of capabilities and $R_\sigma$, in contrast, are structured data that the system executes and filters without any model. The order of the capabilities matters, because later artifacts refer to earlier ones, e.g., an account has to exist before a shell history can belong to it, and a process should be running before a log records its activity. The catalogue holds the fifty-seven templates introduced at the start of Section~\ref{subsec:defender}. They were designed manually for this work, starting from the MITRE Engage description of each activity, so that each one reads like an ordinary event on a server rather than an obvious trap, and each narrative is phrased the way an investigator would summarize what seems to have happened. The complete catalogue is released with the code~\cite{llmhoney_engage}. The three examples below all answer the same activity (i.e., \emph{Lures}), which MITRE links to credential-related techniques such as \emph{Unsecured Credentials} (T1552), \emph{OS Credential Dumping} (T1003), and \emph{Valid Accounts} (T1078). Those examples aim at showing how one activity can be answered with different scenarios:

\begin{itemize}
    \item \emph{``Fake cloud credentials paired with a matching backup log entry and the process that would use them''}. The capabilities create a cloud credentials file, write a log entry showing a backup job that used those credentials, and add the process that runs that job. The scenario fits any role. An attacker looking for secrets finds the credentials first. If it checks whether they are actually in use, the log and the running process confirm it, so a single bait becomes evidence of a real, active workflow.
    
    \item \emph{``A stale administrator account with a weak shell history, a planted SSH key, and a log entry from its last use''}. The capabilities create the account, write a shell history that shows careless habits, plant an SSH key for it, and record a log entry of its last login. The order matters here, because the history and the key must belong to an account that already exists. The scenario also fits any role. An attacker listing users finds a forgotten administrator account whose history and key suggest a way in, while the log explains why nobody has noticed it.
    
    \item \emph{``A decoy cron backup script, the temporary dump file it supposedly produced, a log entry noting the run, and the process that scheduled it''}. The capabilities plant the script, the dump it left behind, the log of its run, and the process that scheduled it. Unlike the other two, this scenario only fits file, database, and web servers, since only a host that stores valuable data would run such a job, so the persona filter of Section~\ref{subsubsec:persona} never offers it on a mail relay or a developer workstation. An attacker looking for data finds the dump together with the job that produced it, which tells it both what the host contains and how it is maintained.
\end{itemize}

The fourth and last step chooses the scenario for this particular attacker. For each reaction, the candidates are the templates that answer one of the activities in $A(\tau)$, that fit the role of the host, and that have not yet been used in the session. A language model then picks one of them given the command the attacker just ran, a description of the identified technique, the machine persona, the capabilities already used, and a short summary of how the attacker has behaved so far (i.e., how fast it works, how widely it is exploring, and whether earlier deceptions caught its attention). The model is asked for the scenario that adds something the attacker has not yet seen while continuing, rather than contradicting, the scenarios already planted in the session. The behaviour summary is built only from what the session itself has revealed, never from the attacker's real profile, so the defensive system adapts to the attacker exactly as a deployed honeypot could. If only one candidate is left, or the model does not return a valid choice, a candidate is drawn at random using the session's seeded generator. Finally, a scenario with fewer than four capabilities is topped up with other capabilities offered by the same activities, so that every reaction deploys up to four artifacts. Four is the length of the longest template in the catalogue, and giving the \textsc{Multi-TTP} control of Section~\ref{subsec:conditions} the same budget means that the two conditions plant the same number of artifacts per reaction and differ in whether those artifacts describe one event. \figurename~\ref{fig:architecture} shows five representative Engage activities beneath the planner.

A narrative on its own does not, however, guarantee that the artifacts fit together, because each artifact is generated separately and could receive an unrelated name. Therefore, consistency is imposed by the system rather than requested from the model. Every artifact in a scenario must carry one shared \emph{subject}, an ordinary service or job name such as \texttt{backup-runner}, \texttt{log-shipper}, or \texttt{cert-renew}, so that a log entry, a running process, and a file on disk visibly refer to the same event. The subject is taken from a fixed list $\mathcal{S}$ of twelve such names, using the session identifier and the scenario narrative:

\begin{equation}
    \text{subj}(\sigma, \text{sid}) = \mathcal{S}\big[\,\text{SHA-256}(\text{sid} \Vert \nu) \bmod |\mathcal{S}|\,\big]
    \label{eq:subject}
\end{equation}

\noindent so a scenario keeps the same subject within a session and receives different subjects in different sessions, and no single name appears on every host. With twelve names, several scenarios of the same session may share a subject, so corroboration can also link artifacts of different scenarios. Once the content of each artifact has been generated, the subject is written into it in two ways. Fields that hold a path keep their directory and extension but take the subject as file name, so \texttt{/var/log/app.log} becomes \texttt{/var/log/backup-runner.log}. Fields that hold a name, such as a process, service, or unit name, take the subject itself. Free text, such as the body of a log, is left exactly as generated, so the artifacts still read naturally. The implications of imposing consistency in this way are discussed in Section~\ref{sec:discussion}.

Table~\ref{tab:scenario} shows a complete example, i.e., a scenario of the \emph{System Activity Monitoring} activity, deployed in response to a persistence technique such as \emph{Scheduled Task/Job} (T1053) on a host whose persona is a database server. The template's three capabilities, topped up with a fourth as described in the fourth step, build the scenario, and the subject \texttt{audit-collect} appears in all of them. An attacker who lists failed services, reads the system logs, inspects running processes, or lists temporary files therefore meets the same name from four different places. Each artifact would be plausible on its own. What the scenario adds is that they point to each other: whichever one the attacker finds first, its name leads to the other three, so following it uncovers the rest of the scenario instead of a dead end.

\begin{table}[H]
    \centering
    \caption{An example of a deployed scenario and the artifacts it plants,
    answering a persistence technique on a host whose persona is a database
    server, with narrative \emph{``an audit or logging service that has visibly
    stopped working, a log capturing the moment it died, and a temporary file
    left mid-write''}. The subject \texttt{audit-collect} is written into every
    artifact, so the artifacts back each other up rather than simply
    appearing side by side.}
    \label{tab:scenario}
    \small
    \begin{tabular}{@{}lll@{}}
        \toprule
        \textbf{Capability} & \textbf{Artifact} & \textbf{Observable through} \\
        \midrule
        Simulate service failure & \texttt{audit-collect.service} & \texttt{systemctl --failed} \\
        Create log file & \texttt{/var/log/audit-collect.log} & \texttt{cat}, \texttt{ls /var/log} \\
        Generate temporal file & \texttt{/tmp/audit-collect.tmp} & \texttt{ls /tmp} \\
        Inject fake process & \texttt{/usr/sbin/audit-collect} & \texttt{ps aux} \\
        \bottomrule
    \end{tabular}
\end{table}

The host must also stay fresh, since an attacker who finds the same artifacts twice concludes that nothing is happening on the host, so repetition is prevented in two ways. First, each session records which scenarios it has already deployed and excludes them from later choices. Second, when a log file already exists, it is rotated to a numbered copy rather than appended to, as \texttt{logrotate} does on a real system, so that even a repeated deployment shows the attacker something it has not seen before.

Together, these steps decide \emph{what} the host should reveal, in response to \emph{what} the attacker is doing, as a scenario that holds together wherever the attacker looks, whereas previous generative honeypots focus on making each individual response look genuine~\cite{sladic2024llm,otal2024llmhoneypot}. The realism the planner aims for is that of a machine with a history, not just that of a convincing shell.

\subsubsection{Artifact Factory}
\label{subsubsec:factory}

Once the planner has chosen a scenario, the artifact factory turns it into real changes on the emulated host. Each capability in the scenario's list is a configurable operation on that host. Twenty-nine capabilities are implemented, such as creating a log file, adding a process to the process table, planting an authorized SSH key, adding a network interface, weakening the password policy, or marking a service as failed. On a production machine, the same operations would correspond to well-known deception techniques: decoy files and logs, honeytokens such as planted credentials or keys, and decoy accounts or services. Here they act on the emulated host, which lets the defender create them instantly and without touching a real system. \figurename~\ref{fig:architecture} shows four of them beneath the factory, fake credentials, a fake process, a planted SSH key, and a fake log, which are the capabilities behind the \emph{Lures} scenarios described in Section~\ref{subsubsec:engage} and cover the places an attacker inspects most often: stored secrets, running processes, access keys, and system logs.

Capabilities do not contain fixed values. Instead, each one declares the fields it needs, and a language model fills those fields with content that fits the scenario being told. Three examples show how this works:
\begin{itemize}
    \item adding a network interface requires a name, a MAC address, and an IPv4 address with its prefix, so on a host that runs containers the model might produce \texttt{docker0} at \texttt{172.17.0.1/16};
    \item creating a log file requires a path and its contents, such as a timestamped line stating that a backup job has finished;
    \item adding a process requires a command line, a process identifier, and resource usage figures, so that the process appears under \texttt{ps aux} alongside the genuine ones, a part of the deception whose realism Section~\ref{subsec:abandonment} examines.
\end{itemize}
Because each capability only declares fields and the model fills them, the same capability can serve any role, but this freedom also has a limit. The system guarantees the form of every artifact: only the declared fields are accepted, names and paths carry the subject of the scenario, the persona filter decides which scenarios are possible at all, and the emulated host stores each artifact in the format of its kind, such as a row of the process table. It does not check the free content against the persona, since the prompt that writes the artifacts receives the scenario but not the list of installed packages, so a log line could, for instance, mention a database that the host does not run. Section~\ref{sec:discussion} returns to this limitation.

All the artifacts of a scenario are always generated together, in a single call to the model that also receives the scenario's narrative and its subject, and the scenario is always deployed as a whole. This characteristic is what allows the artifacts to refer to each other, for example a log entry that names the process that wrote it, and it turns a set of individually plausible artifacts into evidence of one single event.

The factory also keeps track of what each capability has already planted in the session. Once a capability has been used three times, the model also receives the values it produced before and must return something clearly different. Without this rule, a capability used repeatedly would plant almost the same artifact again and again, and the host would look frozen even though the defence keeps reacting. The rule starts at the third use rather than the first because its purpose is to keep a heavily repeated capability from freezing the host, not to change how varied ordinary reactions are, and it applies identically in every condition.

\subsection{Attacking System}
\label{subsec:attacker}

Facing the defensive system is an autonomous attacker that pursues an objective, adapts to what it finds, and decides for itself when a target no longer merits pursuit. Its components appear in the right part of \figurename~\ref{fig:architecture} and form a closed loop, i.e., a profile establishes who the attacker is, a planner chooses what to attempt, actions execute against the host, a judge evaluates what came back, and two explicit rules decide whether the attacker stays, feeding that evaluation into the next planning decision. These components are described next.

\subsubsection{Threat Actor Profile}
\label{subsubsec:profile}

Each session begins by giving the attacker an identity, made of two independent parts, i.e., a threat actor profile, which fixes who the attacker is and what it wants, and a skill level, which decides how capable it is.

The threat actor profile follows the attacker ontology of our previous work~\cite{lopez2026enhancing}. Particularly, it is a record of thirteen attributes: the category of actor, such as criminal, activist, insider, or nation-state; its motivation, such as financial gain, espionage, notoriety, ideology, or sabotage; its attitude, which ranges from opportunistic to targeted, persistent, or destructive; its skills and its prior knowledge of the target; its affiliation; its degree of automation; its typical evasion posture; the kill-chain phase in which it usually operates; a risk tolerance from $1$ to $10$; its preferred target, such as credentials or data; its country; and the threat group it resembles. Eight coherent profiles are defined, ranging from an activist seeking notoriety to a state actor seeking sabotage, and each session draws one of them at random from its seed. The profile shapes both what the attacker pursues and how much doubt it accepts. Its motivation, attitude, skills, knowledge, risk tolerance, and preferred target are given to the attack planner (Section~\ref{subsubsec:planner}) to guide what the attacker pursues, and its attitude and risk tolerance set how much doubt it accepts before concluding that it faces a trap (Section~\ref{subsubsec:judge}). The profile describes who the attacker is and what it wants, whereas how capable it is in the experiment is set only by the skill level described next: the skills attribute of the profile is passed to the planner as part of that description, but it changes neither the techniques available nor the rules that end the session. In particular, the profile has no influence on loss of interest, which depends only on the skill level.

The skill level $\ell \in \{\textsc{Beginner}, \textsc{Intermediate}, \textsc{Advanced}\}$ is an experimental factor, set independently of the profile so that every profile can be tested at every level. Three levels are used because each one corresponds to a different way of judging a target, which Section~\ref{subsubsec:judge} turns into a different rule: a beginner asks whether the host shows any activity at all, an intermediate attacker asks whether it shows enough, and an advanced one asks whether what it shows fits together. A finer scale would split these same three behaviors without adding a new one, and every extra level would add $384$ sessions to the factorial design of Section~\ref{subsec:campaign}. The skill level controls three quantities, as detailed next.

The first quantity is the size of the technique pool. Each session receives a pool of ATT\&CK techniques to work through, drawn from the tactics that suit its level, i.e., six techniques of reconnaissance and discovery for a \textsc{Beginner}; twelve of credential access, collection, execution, and lateral movement for an \textsc{Intermediate}; and twenty of privilege escalation, persistence, defense impairment, command and control, exfiltration, and impact for an \textsc{Advanced} attacker. The tactics follow the progression of an intrusion, so that a beginner stays in the first steps of reconnaissance and discovery while an advanced attacker reaches the steps that require established access, and the pool grows with the level because a more experienced attacker knows more techniques and therefore has more to try on a host. Pools are drawn from the session seed, so repetitions of the same experimental cell receive different pools while remaining reproducible.

The second quantity is the frequency of reconnaissance. Besides its techniques, the attacker periodically runs a fixed reconnaissance sweep, a short set of commands such as \texttt{ps aux}, \texttt{who}, or \texttt{ls -la /tmp/} that checks the current state of the host. The sweep runs after every $\text{interval}(\ell)$ techniques:

\begin{equation}
    \text{interval}(\ell) =
    \begin{cases}
        3 & \ell = \textsc{Beginner} \\
        2 & \ell = \textsc{Intermediate} \\
        1 & \ell = \textsc{Advanced}
    \end{cases}
    \label{eq:recon}
\end{equation}

\noindent An experienced attacker re-checks the host more often because it watches for change, such as a new user, process, or log entry, which can reveal both a defender and a fresh opportunity, whereas a novice follows its plan and looks back less. A more capable attacker therefore has more chances to notice what the defender plants, while the responses to the sweep do not count towards detection (Section~\ref{subsubsec:judge}).

The third quantity is how long the attacker keeps going when the host stops showing anything new. Its base patience, written $\beta(\ell)$, is the number of consecutive rounds without new findings that it accepts before leaving, and it takes the same values as Equation~\ref{eq:recon}: three rounds for a \textsc{Beginner}, two for an \textsc{Intermediate}, and one for an \textsc{Advanced} attacker. An expert has seen many real servers and could recognize an empty one sooner, whereas a novice keeps trying for longer. This short base patience is what makes the advanced attacker the hardest one to retain, which matters for the results of Section~\ref{sec:results}. On top of it, the attacker earns extra patience when the host rewards it, and the skill level also determines \emph{what kind} of reward earns that patience, as Section~\ref{subsubsec:judge} explains. That extra patience has the same ceiling at every level, so an expert never receives more of it than a novice: it only starts from less and responds to a different kind of reward.

\subsubsection{Attack Planner}
\label{subsubsec:planner}

After each technique, the attacker decides what to do next. That decision is taken by a language model that, like the defender, reasons over the history of the session. Writing $O_t$ for the outcomes of the techniques attempted up to round $t$, where a round is one technique together with its follow-up commands, the planner implements:

\begin{equation}
    \Pi: \big(O_t,\; \mathcal{T}_{\text{rem}},\; p_{\text{actor}}\big) \mapsto \tau_{t+1} \in \mathcal{T}_{\text{rem}} \cup \{\bot\}
    \label{eq:planner}
\end{equation}

\noindent where $\Pi$ is the planning function realized by the language model, $\mathcal{T}_{\text{rem}}$ is the part of the session's technique pool not yet attempted, $p_{\text{actor}}$ is the threat actor profile of Section~\ref{subsubsec:profile}, $\tau_{t+1}$ is the next technique to execute, and $\bot$ means that no further technique will be attempted.

Each outcome in $O_t$ summarizes one technique: which technique it was, how many commands it ran, the mean credibility of the responses it received, whether any of them looked suspicious, and how much new content it uncovered, of which kinds, and how much of it corroborated earlier findings. For example, an outcome may record that \emph{Unsecured Credentials} (T1552) ran four commands with a mean credibility of $0.95$, that is, the average of the scores assigned by the Credibility Judge of Section~\ref{subsubsec:judge}, raised no suspicion, and uncovered seven new items of two kinds, a credentials file and a log entry referring to it. The planner sees how each technique went, namely its commands, its credibility, and whether it looked suspicious, together with the profile, a summary of the session, and the techniques still available. A history that looks genuine leads it to keep working the target rather than switching to quieter or more cautious techniques, especially if its profile is persistent or targeted. What a technique uncovered, on the other hand, is not shown to the planner but feeds the rules of Section~\ref{subsubsec:judge}, which decide whether the attacker is still interested in the host.

The planner reasons only over what the attacker has seen, namely the output of its commands and the credibility scores the judge assigned to them, and it is never told which experimental condition is active, nor whether the defensive system reacted to anything it did. It also cannot end the session on its own while techniques remain, and must always pick the next one, because a language model left free to decide when it is done tends to declare itself satisfied at arbitrary points, which would make the length of a session depend on the model rather than on what the host showed. A session therefore ends in only three ways, and no time limit or fixed number of steps cuts it short:
\begin{itemize}
    \item \emph{detection}, when the attacker concludes that the host is a trap;
    \item \emph{loss of interest}, when the host has stopped offering anything new;
    \item \emph{exhaustion}, when the attacker runs out of techniques.
\end{itemize}
The first two are decided by the explicit rules of Section~\ref{subsubsec:judge}.

Two mechanisms let a rewarding host keep the attacker for longer, so that engagement can show up as session length instead of being capped by design, and neither of them knows which experimental condition is active:
\begin{itemize}
    \item \emph{reserve techniques}: when the attacker exhausts its assigned pool and the session so far has looked convincing, the planner may ask for an extra technique instead of stopping. Extra techniques come from a small reserve of two, four, or six techniques depending on the skill level, revealed one at a time, and this is the only case in which the planner returns $\bot$, namely when the pool is empty and it declines to ask for more;
    \item \emph{follow-up commands}: when a result uncovers at least five new items, the attacker may run up to three short commands to inspect exactly what it has just found before moving on, such as reading a file that has just been mentioned, checking a process identifier that has just appeared, or looking up a user that has just been seen, as a real attacker investigates a discovery on the spot instead of waiting for its next planned technique.
\end{itemize}

\subsubsection{Actions on the Victim Machine}
\label{subsubsec:actions}

This component is where the attacker acts, i.e., it turns each technique chosen by the planner into commands, types them into the SSH session one at a time, and reads what the host answers, mimicking a human attacker at a terminal, so the attacker never describes what it intends to do. Consequently, anything the defender has planted is put to the same test a real attacker would apply, since it is found, or missed, only through the commands the attacker actually runs. The component also turns each answer into the observations that the rest of the attacker relies on, namely, the credibility score given by the judge and the list of things the answer revealed.

The commands are taken from Atomic Red Team~\cite{atomicredteam}, an open catalogue of small tests that reproduce ATT\&CK techniques with real commands. Relying on a public catalogue, rather than letting a language model write the commands, keeps the attack reproducible, ties every action to documented adversary behavior, and ensures that a given technique is always exercised by the same commands, whichever model plays the attacker. Atomic Red Team is the source used in this work rather than a requirement of the design. The component only needs, for each technique, a list of commands to send, so the same architecture can take them from another public catalogue, from commands recorded in real intrusions, such as the logs of a production honeypot, or from a language model when variety matters more than reproducibility, without changes to the defensive system or to the rules that end a session. Each technique offers one or more tests, and each test is written for one or more operating systems. The component keeps only the tests written for the operating system the honeypot presents. In our use case, this is Linux, since the environment emulates an Ubuntu server; a honeypot emulating another system would simply keep that system's tests, with no change to the rest of the design.

As for which parts of an attack are covered, the techniques come from the ATT\&CK tactics assigned to each skill level in Section~\ref{subsubsec:profile}, which cover the attack from the moment the attacker is inside the host: learning about the host, stealing credentials, running code, gaining privileges, staying on the host, talking to the outside, extracting data, and causing damage. Resource development and initial access are deliberately left out, because they happen before the attacker reaches the host or away from it, such as renting infrastructure or sending a phishing e-mail. In this work, every session starts once the attacker is already logged in, so the honeypot could neither observe those steps nor answer them. \figurename~\ref{fig:architecture} summarizes the attack in four representative phases, i.e., reconnaissance, credential access, privilege escalation, and discovery. These are the phases that an attacker who has just obtained a shell on a Linux server typically goes through first, whatever its final goal, and they are also the phases in which it reads most of the host's content, which is where planted deception is found.

Before anything is sent, each test is turned into the commands a terminal would receive. Its placeholders take the default values of the catalogue, so \texttt{ls -a >> \#\{output\_file\}} becomes \texttt{ls -a >> /tmp/T1083.txt}; lines split with a trailing backslash are joined and comment lines are dropped; commands that would close the connection, such as \texttt{exit} or \texttt{reboot}, are skipped, so that only the attacker's own judgment ends a session; and the remaining commands are sent one at a time, each after the previous output has arrived. Cleanup commands are not run, since every session starts from a fresh copy of the host (Section~\ref{subsubsec:environment}), and as soon as a response is judged suspicious the rest of that test is abandoned, as an attacker who sees something odd would do.

Once a command has run, its output goes to the credibility judge of Section~\ref{subsubsec:judge}, which scores how genuine it looks. In addition, the component extracts from the output the identifiers an attacker would remember, such as file names, paths, user names, process names, and addresses, and keeps those it has not seen before. That is, these identifiers are the discoveries counted by the interest rules of Section~\ref{subsubsec:judge}. Before the first technique runs, the component also performs one reconnaissance sweep, with the same commands as the periodic sweep of Section~\ref{subsubsec:profile}, and records everything the host shows on arrival. Everything this first sweep shows is treated as already seen, so it never counts as a discovery: only what appears afterwards, such as the artifacts the defensive system plants, can show the attacker that the host has something worth the effort.

Because the environment keeps state, however, the attacker's own actions show up in later answers. An attacker that creates \texttt{/tmp/T1083.txt} and later lists \texttt{/tmp} would otherwise count its own file as a sign that the host is rich in content, whereas a real attacker is never surprised by its own traces. The component therefore counts as a \emph{discovery} only an identifier that tells the attacker something new about this host, and discards three kinds. First, identifiers the attacker has typed itself in any earlier command, including every trailing part of a typed path, because the shell often repeats only part of what was typed: running \texttt{/tmp/t1040\_linux\_pcapdemo}, for instance, fails with \texttt{bash: t1040\_linux\_pcapdemo: command not found}, which repeats the file name but not its directory. Second, identifiers the attacker had already seen, either in the sweep on arrival or in an earlier answer. Third, identifiers that say nothing about this particular host: generic shell vocabulary, the stock content every Ubuntu server ships with, and values that change on every call, such as timestamps and process identifiers. Identifiers are compared in lower case, so a name is recognized whatever its capitalization, and the complete lists are released with the code~\cite{llmhoney_engage}. In what follows, $D_j$ denotes the discoveries of command $c_j$.

To illustrate the whole process, suppose that a \textsc{Beginner} attacker is given \emph{File and Directory Discovery} (T1083). The component keeps the three tests of this technique written for Linux, replaces the placeholder of the first one with \texttt{/tmp/T1083.txt}, and runs commands such as \texttt{ls -a >> /tmp/T1083.txt} and \texttt{find . -type f}. The judge scores each output, and the identifiers the attacker typed, among them \texttt{t1083.txt}, are remembered as its own. Later, the periodic sweep runs \texttt{ls -la /tmp/}. Its answer lists \texttt{T1083.txt}, which is ignored because the attacker created it, and, if the defender has reacted in the meantime, a file such as \texttt{backup-runner.lock} that was not there on arrival. Only that file counts as a discovery, and if a log or a process carrying the same name turns up later, the two findings corroborate each other (Equation~\ref{eq:corroboration}).

\subsubsection{Credibility Judge}
\label{subsubsec:judge}

The last component decides whether the attacker stays. As described in Section~\ref{subsubsec:planner}, a session ends in one of three ways: the attacker concludes that the host is a trap, concludes that the host is not worth the effort, or runs out of techniques. The last one needs no decision, and the component applies one explicit rule for each of the other two, called detection and loss of interest and described in turn below. Both rules are fixed formulas rather than decisions of a language model, so that the moment at which a session ends depends on what the host showed and not on how the model playing the attacker happens to behave.

Detection starts with a language model, the judge, which evaluates every response the host returns, following the LLM-as-a-judge paradigm~\cite{zheng2023judging}. The judge receives the command and its output and assigns a credibility score $s \in [0,1]$, expressing how convincingly the response resembles what a genuine server of the kind the honeypot presents would return. A response scoring below a threshold $\theta = 0.4$ is marked as suspicious, and a technique is suspicious when at least one of the responses to its own commands, or to its follow-up commands, is. The responses to the periodic reconnaissance sweep of Section~\ref{subsubsec:profile} do not count towards detection, and they feed only the loss-of-interest rule described below, since the sweep is a routine glance that repeats the same commands at fixed intervals, whereas suspicion is judged on what the attacker actively probes with each technique. Because the judge is itself a language model, detection in this work means detection by this judge under this rule, and Section~\ref{subsec:realism} checks the realism of the host by two means that do not depend on it.

Section~\ref{subsec:abandonment} examines how the results change when the sweep also counts. Failed commands receive no special treatment: a missing binary or a denied permission is answered as a real server would answer it, and the judge scores that answer like any other. Should the call to the judge itself fail, the response is given the benefit of the doubt and scored as fully credible, so a technical fault on the attacker's side can never push it towards leaving. How much suspicion an attacker tolerates before writing a host off depends on its threat actor profile and on its skill level. Writing $n_{\text{s}}(t)$ for the number of suspicious techniques up to round $t$, the attacker concludes that the host is a trap when

\begin{equation}
    n_{\text{s}}(t) \;\ge\; \max\Big(2,\; \big\lfloor \psi(p_{\text{actor}}) \cdot \delta(\ell) \big\rceil \Big)
    \label{eq:detection}
\end{equation}

\noindent where $\lfloor\cdot\rceil$ denotes rounding to the nearest integer, $\psi(p_{\text{actor}})$ is the suspicion tolerance of the actor, that is, the number of suspicious techniques it accepts, not to be confused with the patience of the loss-of-interest rule below, and $\delta(\ell)$ scales that tolerance by skill level with the values of Table~\ref{tab:attacker_params}. Suspicion tolerance derives from the attitude in the profile: $1$ for opportunistic actors, which leave at the first solid sign, $2$ for destructive ones, and $3$ for targeted or persistent actors, which came for this specific host and want confirmation, with one more for profiles whose risk tolerance is $7$ or higher. The skill scaling makes a capable attacker need less evidence, with $\delta < 1$, and a novice more, with $\delta > 1$. The lower bound of two is a design choice of this work: a single suspicious technique is not yet a pattern, since one low score can come from a harsh judgment or from a command that legitimately fails, as a missing binary does on any real server, whereas two suspicious techniques mean that the anomaly has repeated. The bound only matters for the least tolerant actors, the opportunistic profiles at \textsc{Intermediate} and \textsc{Advanced} level, for which $\psi \cdot \delta$ rounds to one, and Section~\ref{subsec:abandonment} shows that lowering it to one would change the outcome of very few sessions.

\begin{table}[H]
    \centering
    \caption{Attacker parameters by skill level. The pool, the reserve, the reconnaissance interval, and $\beta(\ell)$ are introduced in Section~\ref{subsubsec:profile}, $\delta(\ell)$ in Equation~\ref{eq:detection}, and the ceiling of the extra patience $\gamma_\ell$ in Equation~\ref{eq:gamma}.}
    \label{tab:attacker_params}
    \small
    \begin{tabular}{@{}lccc@{}}
        \toprule
        \textbf{Parameter} & \textsc{Beg.} & \textsc{Int.} & \textsc{Adv.} \\
        \midrule
        Technique pool size & 6 & 12 & 20 \\
        Reserve techniques & 2 & 4 & 6 \\
        Recon interval (Eq.~\ref{eq:recon}) & 3 & 2 & 1 \\
        Base patience $\beta(\ell)$ & 3 & 2 & 1 \\
        Detection scaling $\delta(\ell)$ (Eq.~\ref{eq:detection}) & 1.5 & 1.0 & 0.85 \\
        Ceiling of extra patience $\gamma_\ell$ (Eq.~\ref{eq:gamma}) & 12 & 12 & 12 \\
        \bottomrule
    \end{tabular}
\end{table}

Nevertheless, a host can be entirely convincing and still offer nothing worth the effort, an outcome the defensive system must equally avoid. Thus, the second rule looks at what the attacker discovers rather than at the credibility scores. Loss of interest, as modelled here, means that the host has stopped offering anything valuable for the attacker. It covers both an unmotivated attacker that simply loses interest and a motivated, skilled one that prefers to move on to a more promising target, since the patience of the attacker depends on its skill level rather than on the motivation in its profile, and telling the two apart would require tying patience to motivation, which is left as future work. Discoveries are the identifiers defined in Section~\ref{subsubsec:actions}.

The rule works in rounds, where a round is one technique together with its follow-up and reconnaissance commands, and the new discoveries of round $i$, written $N_i$, gather the sets $D_j$ of all its commands. Also, each discovery is assigned a kind of artifact according to its shape, such as a log, a process, a credential, a network interface, an account, a package, a backup, or a configuration file, which gives for each round $i$ the set $K_i$ of kinds it revealed. A round is productive when it reveals at least $\eta = 5$ new identifiers. A threshold is needed because the host's own output varies slightly from one command to the next and yields a trickle of new identifiers even when nothing has been planted. Its value was set from the pilot sessions described below, in which the median round revealed four new identifiers when the defender reacted with a single MITRE Engage activity and six when it combined several, whereas a round in which a scenario was deployed revealed between $10$ and $33$. A threshold of five stops the background trickle from counting while any substantial deployment still does.

The attacker loses interest when it has gone too long without a productive round:

\begin{equation}
    \text{abandon}(t) = \Big[\, \text{streak}(t) \;\ge\; \beta(\ell) + \gamma_\ell(O_t) \,\Big],
    \qquad
    \text{streak}(t) = \sum_{i = t_0}^{t} w_i,
    \qquad
    w_i = \begin{cases} 2 & \text{fam}(i) = \text{fam}(i-1) \\ 1 & \text{otherwise} \end{cases}
    \label{eq:abandon}
\end{equation}

\noindent where $t_0$ is the first round of the current run of unproductive rounds, so that $\text{streak}(t)$ counts the consecutive unproductive rounds up to round $t$, each with weight $w_i$, $\text{fam}(i)$ is the ATT\&CK family of the technique of round $i$, $\beta(\ell)$ is the base patience of Table~\ref{tab:attacker_params}, and $\gamma_\ell(O_t)$ is the extra patience that the attacker has earned from what the host showed earlier. A round in the streak counts twice when its technique belongs to the same ATT\&CK family as the previous one, such as \texttt{T1053.002} followed by \texttt{T1053.003}, because repeating the same kind of move with nothing to show for it is a stronger sign of a dead end than an equally long run of varied attempts. In other words, the attacker accepts a dry stretch only if the host has earlier proved worth it, and $\gamma_\ell$ measures how long a stretch it accepts. That extra patience is capped at $12$ at every level, and the skill level decides what kind of reward earns it:

\begin{equation}
    \gamma_\ell(O_t) = \min\big(12,\; g_\ell(O_t)\big), \qquad
    g_\ell(O_t) =
    \begin{cases}
        4\min(\rho_t, 2) + \rho_t & \ell = \textsc{Beginner} \\[2pt]
        b(\kappa_t) + \left\lfloor \pi_t / 2 \right\rfloor & \ell = \textsc{Intermediate} \\[2pt]
        2\left\lfloor \chi_t / 3 \right\rfloor + \min(\kappa_t, 3) & \ell = \textsc{Advanced}
    \end{cases}
    \label{eq:gamma}
\end{equation}

\noindent where $\rho_t$ is the number of rounds that revealed at least one kind of artifact, $\kappa_t = |\bigcup_{i \le t} K_i|$ is the number of different kinds of artifact seen so far, $\pi_t$ is the number of productive rounds, $\chi_t$ is the corroboration defined in Equation~\ref{eq:corroboration} below, and $b(\kappa) = \min(\kappa, 4) + \lfloor (\kappa - 4)/2 \rfloor$, applied only when $\kappa > 4$ and equal to $\kappa$ otherwise, counts every kind beyond the fourth at half value, since the first few kinds of artifact are what persuade an attacker that the host is alive whereas the tenth adds much less than the third.

Read in rounds of patience, the constants of Equation~\ref{eq:gamma} are as follows. A \textsc{Beginner} earns four rounds for each of the first two rounds that reveal anything, and one for each later one. An \textsc{Intermediate} attacker earns one round for each kind of artifact it has seen, half a round for each kind beyond the fourth, and one more for every two productive rounds. An \textsc{Advanced} attacker earns two rounds for every three discoveries that corroborate earlier ones, and at most three rounds from variety. These constants are relative weights chosen when the rule was designed, following the question each level asks, rather than measured quantities, and Section~\ref{subsec:abandonment} shows how the results change when the weight of corroboration is reduced or removed.

For example, an \textsc{Advanced} attacker has a base patience of one round. If the host has shown it eight kinds of artifact that share no names, then $\chi_t = 0$ and $\gamma_\ell = 0 + \min(8,3) = 3$, so the attacker leaves after four rounds without new findings. If the same artifacts instead share names and $\chi_t = 12$, then $\gamma_\ell = 2 \cdot 4 + 3 = 11$, and the attacker accepts twelve such rounds. A \textsc{Beginner} that has seen two rounds with some artifact earns $4 \cdot 2 + 2 = 10$ and accepts thirteen, whereas against a host that never reacts, where $\rho_t = 0$, it leaves after its base three. An \textsc{Intermediate} attacker that has seen six kinds of artifact over four productive rounds earns $b(6) + 2 = 5 + 2 = 7$ and accepts nine.

Each branch of Equation~\ref{eq:gamma} therefore answers a different question: a novice asks whether there is any activity at all, an intermediate attacker whether there is a lot of it, and an experienced one whether its findings fit together, which is why variety contributes at most three rounds to its patience while corroboration counts double. Recognizing that separate findings fit a single underlying event is the kind of judgment an experienced investigator makes and an inexperienced one does not. That an experienced attacker values corroboration is therefore an assumption of the attacker model, motivated by how investigators work, and not a result of this study; Section~\ref{subsec:abandonment} reports which conclusions depend on it.

Corroboration measures whether the discoveries describe a single event, by counting how many of them share their name with another, different discovery:

\begin{equation}
    \chi_t = \sum_{i \le t} \Big| \big\{\, x \in N_i \;:\; |\{ y \in V_i : \text{stem}(y) = \text{stem}(x) \}| \ge 2 \,\big\} \Big|
    \label{eq:corroboration}
\end{equation}

\noindent where $N_i$ are the new discoveries of round $i$, $V_i$ is everything the attacker has seen up to that round, including what the host showed on arrival, and $\text{stem}(x)$ is the bare name of $x$, that is, what remains after removing its directory and its extension, so that \texttt{/var/log/backup-runner.log} and \texttt{/usr/sbin/backup-runner} share the stem \texttt{backup-runner}. Each discovery therefore counts once, when it is made, if by then at least one other discovery carries the same name. The attacker cannot tell whether a discovery was planted by the defender or belonged to the host from the start, so all of them are considered. A log, a process, and a binary with the same name corroborate each other, whereas a pile of unrelated artifacts, each with its own name, scores zero. Numbered copies of a rotated file, such as \texttt{app.log.1}, count as a single file, since they repeat one artifact rather than add to the scenario, and names shorter than five characters are ignored because short names coincide by chance too often. The divisor of three in Equation~\ref{eq:gamma} turns corroboration into patience at a rate such that a scenario, whose $3.44$ capabilities on average corroborate one another several times as the session goes on, earns enough patience to survive the quiet stretch that follows once all its artifacts have been found.

This patience model is a proposal of this work, and it has three properties that make it suitable for measuring deception. First, it is interpretable, since every term corresponds to something the attacker has observed and could explain, such as how many kinds of artifact it has seen or how many of them share a name. Second, it is blind to the experimental condition, since it reads only the attacker's own observations, so the same rule applies identically whatever the defensive system does. Third, as a fixed rule rather than a decision of the language model, it behaves the same whichever model plays the attacker, so differences between conditions cannot come from one model simply losing interest sooner than another. The structure of Equation~\ref{eq:gamma}, that is, which signal each level responds to and the order of their weights, follows from the threat actor model of Section~\ref{subsubsec:profile}: a novice is rewarded by any activity, an intermediate attacker by variety, and an expert mostly by corroboration. The numeric values were then fixed in a pilot campaign of $36$ sessions per configuration under two constraints: the ceiling is the same at every level, and no single kind of reward can fill the whole allowance on its own, since otherwise a large enough deployment of any kind would hold every attacker whatever it showed. Other values satisfying the same constraints would serve equally well, so the values reported here are one valid choice rather than a unique optimum. Because the rule only decides when a session ends, any other choice that never grants more patience can be applied exactly to the recorded sessions, and Section~\ref{subsec:abandonment} uses this to show how the results depend on it, while Section~\ref{subsec:generalisation} shows that the behavior holds across models that took no part in the pilot. Table~\ref{tab:attacker_params} gathers the numeric attacker parameters that depend on the skill level.


\section{Experimental Setup}
\label{sec:setup}

This section describes how the system presented in Section~\ref{sec:design} was evaluated. In particular, Section~\ref{subsec:conditions} defines the ablation conditions used to separate the contribution of the main design elements, whereas Section~\ref{subsec:models} specifies the models and infrastructure on which the campaign ran. Section~\ref{subsec:campaign} then details the factorial design and its execution, and Section~\ref{subsec:metrics} sets out the measurements recorded for every session.

\subsection{Ablation conditions}
\label{subsec:conditions}

Assessing whether narrative coherence contributes anything beyond the mere presence of deception requires more than comparing the proposed system against a honeypot that never reacts~\cite{provos2004honeyd,cowrie}. Such a comparison would confirm only that reacting beats not reacting, a result recent work on generative honeypots already establishes~\cite{sladic2024llm,otal2024llmhoneypot}, leaving open whether the reaction needs to be coherent or merely plentiful. Four conditions therefore participate in the evaluation, obtained by progressively removing design elements from the deception engine described in Section~\ref{subsec:defender} while holding every other component fixed. Each step between consecutive conditions changes one aspect of the deception, namely whether it reacts, how varied it is, and whether its artifacts form a scenario, but each step is a change between two complete designs rather than the removal of a single variable, so the difference between consecutive conditions is attributed to that aspect as a whole.

\begin{description}
    \item[\textsc{Static}.] Reactive deception is disabled entirely. The honeypot behaves like a conventional emulated honeypot: it answers every command from the host defined by its machine persona, but never plants anything in response to the attacker. Whatever the attacker does, the host keeps showing the content it showed on arrival, apart from the traces the attacker leaves itself. For example, an attacker that lists \texttt{/var/log/} after trying to harvest credentials finds exactly the log files that were there when it logged in, and repeating the command later returns the same answer. This condition establishes the floor of the comparison: any engagement it produces is due to the emulated host alone, so whatever the other conditions add above it can be attributed to deception.

    \item[\textsc{Same-TTP}.] The defender reacts, deploying several capabilities per reaction, up to the same number as the complete system, but all of them come from a single MITRE Engage activity, chosen at the first reaction and kept for the whole session. The attacker therefore meets deception repeatedly, but always of the same kind, for instance more and more log files. Keeping the activity fixed is what defines this condition, i.e., if a new activity were chosen at every reaction, successive reactions would show different kinds of artifact, and the condition would become the next one.

    \item[\textsc{Multi-TTP}.] Capabilities are drawn at random from all the activities that match the identified technique, up to the same number per reaction as the complete system, and the content of each one is generated separately, without any scenario. The attacker thus meets the same budget of artifacts per reaction as under the complete system, of a similar variety, while those artifacts bear no relation to one another, such as the log of one service, the binary of another, and the unit of a third. This condition is the critical control. What separates it from the complete system is scenario coherence as a whole: the curated template, the ordered capabilities, the shared subject, the joint generation of all the artifacts, and a planner that chooses with a summary of the attacker's behavior.

    \item[\textsc{Chain}.] The complete system of Section~\ref{subsec:defender}. Each reaction deploys a curated scenario whose capabilities describe a single event and whose artifacts share a common subject, so that, for instance, the log, the binary, and the service of the same backup job appear together. This is the condition on which the proposal stands, and coherence is expected to pay off for three reasons. On a genuine server, traces are connected because they share a cause, whereas unrelated artifacts pointing nowhere are what an experienced attacker expects from a trap. Every finding becomes a lead, since a log that names a process invites the attacker to look for that process, so the scenario itself invites further commands. And artifacts that confirm one another pass the cross-check by which an investigator separates real evidence from fabricated evidence. Since \textsc{Chain} deploys the same budget per reaction as \textsc{Multi-TTP}, whatever it gains comes from how the material is arranged rather than from its amount.
\end{description}

Table~\ref{tab:conditions} illustrates with an example what the attacker could observe after one reaction under each condition, which is the only information available to it. Ordered as $\textsc{Static} \prec \textsc{Same-TTP} \prec \textsc{Multi-TTP} \prec \textsc{Chain}$, the conditions form a \emph{deception ladder}, in which each condition adds one aspect of the deception to the condition below, and each step up answers a separate question with its own practical meaning.

\begin{table}[H]
    \centering
    \caption{Example of what the attacker could observe after one reaction under each condition. The last two rows plant the same number of artifacts spanning the same number of kinds, and differ in whether those artifacts share a subject and describe one event.}
    \label{tab:conditions}
    \small
    \begin{tabular}{@{}llcc@{}}
        \toprule
        \textbf{Condition} & \textbf{Artifacts observed} & \textbf{Kinds} & \textbf{Naming} \\
        \midrule
        \textsc{Static} & --- & 0 & --- \\
        \addlinespace[2pt]
        \textsc{Same-TTP} & \texttt{nginx-access.log} & 1 & unrelated \\
                          & \texttt{apache-error.log} & & \\
                          & \texttt{mysql-slow.log} & & \\
        \addlinespace[2pt]
        \textsc{Multi-TTP} & \texttt{redis-server.log} & 3 & unrelated \\
                           & \texttt{/usr/sbin/ntpd} & & \\
                           & \texttt{squid.service} & & \\
        \addlinespace[2pt]
        \textsc{Chain} & \texttt{backup-runner.log} & 3 & \textbf{shared} \\
                       & \texttt{/usr/sbin/backup-runner} & & \\
                       & \texttt{backup-runner.service} & & \\
        \bottomrule
    \end{tabular}
\end{table}

The first step, from \textsc{Static} to \textsc{Same-TTP}, asks whether reacting to the attacker is worth it at all. It corresponds to moving from a conventional honeypot, which offers the same content to every visitor, to one that plants lures in response to what each attacker does. Reacting is expected to retain the attacker for longer at every skill level, and above all a \textsc{Beginner}, who only needs to see that the host is active.

The second step, from \textsc{Same-TTP} to \textsc{Multi-TTP}, asks whether variety matters, keeping the same budget of artifacts per reaction while widening the range of defensive activities. It corresponds to deploying decoys of many kinds rather than many decoys of one kind. Variety is expected to help most against an \textsc{Intermediate} attacker, who judges a host by how much it has to offer.

The third step, from \textsc{Multi-TTP} to \textsc{Chain}, asks whether coherence matters, keeping the same budget per reaction and a similar variety, and it carries the central claim of this work. Its practical meaning is direct, i.e., a defender with a given budget of artifacts can either plant them independently or arrange them into a scenario, and this step measures what the arrangement is worth. Coherence is expected to make the largest difference against an \textsc{Advanced} attacker, who is the hardest to retain, the most valuable to observe, and the most likely to cross-check what it finds. If the proposal is right, \textsc{Chain} should retain such an attacker where \textsc{Multi-TTP}, with the same material, does not.

It is important to remark that every condition shares the same attacker, the same emulated environment, the same persona mechanism, and the same models. Critically, the attacking system receives no indication of which condition is active, as established in Section~\ref{subsec:architecture}, so its behavior can respond only to what the target actually revealed.

\subsection{Models and infrastructure}
\label{subsec:models}

Four open-weight language models were evaluated: \texttt{llama3.1}~\cite{dubey2024llama3}, \texttt{qwen2.5}~\cite{qwen2025}, \texttt{mistral}~\cite{jiang2023mistral}, and \texttt{WhiteRabbitNeo-2-8B}~\cite{whiterabbitneo}, a variant of \texttt{llama3.1} fine-tuned on offensive security material. They were chosen according to three criteria. First, they have open weights and run locally, which a deployed honeypot requires, since the commands of an attacker should not be sent to an external service, and which makes the experiments reproducible without depending on the availability or the silent updates of a commercial service. Second, all of them have between seven and eight billion parameters, so that differences between them reflect how they were trained rather than their size, and all of them fit on a single workstation, as a honeypot defender would be deployed. Third, they cover three independent families of general-purpose models, from three different developers, plus one model specialised in offensive security, which tests whether training on security material changes the result for either side.

The number of models is a trade-off between generality and cost. Since every model plays both roles, the number of model pairs grows with the square of the number of models, and each pair requires $72$ sessions, one per combination of condition, skill level, persona mode, and repetition (Section~\ref{subsec:campaign}). Four models give $16$ pairs and $1152$ sessions, whereas a fifth model would raise the campaign to $1800$ sessions and a sixth to $2592$, and three models would reduce it to $648$ but would leave out either the security-specialized model or one of the families. Four is therefore the smallest set that meets the three criteria while supporting the cross-model analysis of Section~\ref{subsec:generalisation}, in which twelve of the sixteen pairs run a different model on each side. Larger or proprietary models were deliberately left out, since a defender relying on them could not run locally, and whether they would strengthen the effect is left as future work.

All experiments ran on a single workstation built around an NVIDIA GB10 Grace Blackwell superchip, pairing a twenty-core ARM CPU, ten Cortex-X925 and ten Cortex-A725 cores, with a Blackwell GPU of compute capability $12.1$ over $121$~GB of unified memory shared between both processors. The host runs Ubuntu $24.04$ LTS on kernel $6.17$ with Docker $29.2$. We believe that reporting that configuration matters because the latencies of Section~\ref{subsec:cost} are properties of this deployment rather than of the method. In fact, inference on unoptimized local serving differs by an order of magnitude from the tuned deployments that published benchmarks measure~\cite{mlperf2025}. Nonetheless, every experimental condition shares the same machine, so hardware cannot account for any difference observed between them.

All models run locally on the same workstation through Ollama~\cite{ollama}, an open-source server for language models, which both agents query in the same way. The campaign proceeds one model pair at a time: the weights of the two models are loaded, all the sessions of that pair are run, and the weights are removed before moving on to the next pair.

The temperature of each role is a design choice of this work, set according to what the role needs. Temperature controls how much randomness a model adds when it writes its answer: at $0$ it always gives its most likely answer, and higher values make its answers more varied. The technique correlator runs at $0$, because mapping a command to an ATT\&CK technique has a correct answer, and any randomness would only add noise to every later decision of the defender. The Engage planner and the artifact factory run at $0.5$, because their purpose is to produce varied content: at $0$ they would plant the same artifacts in every session, which an attacker visiting several honeypots would soon recognize. Higher values were avoided, since raising the temperature makes generated text only slightly more novel while making it noticeably less coherent~\cite{peeperkorn2024temperature}, and coherence is what the proposal relies on. The attacking roles, namely the Attack Planner, including its choice of follow-up commands, and the Credibility Judge, run at $0.2$, low enough for their decisions to be stable and reproducible, yet above $0$ so that repetitions of the same configuration do not follow exactly the same path. Keeping these temperatures low costs nothing in quality, since for reasoning tasks temperatures between $0$ and $1$ have no significant effect on accuracy~\cite{renze2024temperature}. Finally, the credibility judge uses \texttt{llama3.1} in every configuration, whichever models play the defender and the attacker, so that credibility scores remain comparable across the whole campaign.

\subsection{Campaign design}
\label{subsec:campaign}

The evaluation follows a full factorial design over five factors: the defender's model and the attacker's model, with four values each, the deception condition, with the four values of Section~\ref{subsec:conditions}, the attacker's skill level, with three, and the machine persona, either active or disabled. Every combination of factors is run three times, which gives

\begin{equation}
    N = \underbrace{4 \times 4}_{\text{model pairs}} \times \underbrace{4}_{\text{conditions}} \times \underbrace{3}_{\text{skill}} \times \underbrace{2}_{\text{persona}} \times \underbrace{3}_{\text{repetitions}} = 1152
    \label{eq:sessions}
\end{equation}

\noindent sessions in total. To our knowledge, this makes the campaign one of the largest controlled evaluations of LLM-driven deception, an order of magnitude beyond evaluations based on human participants, such as the $89$ attackers who assessed shelLM~\cite{sladic2024llm}. Running both sides automatically is what makes this scale possible, and the full factorial design means that every factor is tested in combination with every other, so that no conclusion depends on one particular model, skill level, or persona.

Each combination of factors is run three times. The comparisons that matter most are between conditions at a given skill level, and each of them rests on $96$ sessions, from $16$ model pairs, $2$ persona modes, and $3$ repetitions. With $96$ sessions, the abandonment rate of a condition is known to within about ten percentage points at $95\%$ confidence, which is enough to separate conditions whose difference matters in practice. Two repetitions would widen this margin to about twelve points, while a fourth would add $384$ sessions to narrow it by about one.

The persona factor tests a second, smaller hypothesis: that the host is more convincing when it looks like a specific kind of server. With the persona active, each session draws one of the six roles of Section~\ref{subsubsec:persona}, such as a web or a database server, and the host shows the packages, services, and scenarios of that role. With the persona disabled, the host is a bare Ubuntu server with no particular role. Comparing the two separates the value of a consistent host identity from the value of the reactive deception.

Sessions run one at a time, because several sessions running at once would share the same GPU, every model would answer more slowly, and the attacker could take slow answers as a sign that the host is artificial, which would distort the very credibility measurements the study relies on. Each session also starts from scratch, with a fresh copy of the host, a newly drawn machine persona and threat actor profile, and a newly drawn technique pool, so nothing carries over from one session to the next. These random choices are drawn from a seed specific to each session. Running a session again with the same seed therefore draws the same persona, profile, and technique pool, and since every session has its own seed, the choices made for one session do not influence those of any other.

No time limit constrains a session, and this choice is central to the realism of the evaluation. A real attacker is not bound by a fixed session length: it keeps working a host for as long as the host rewards it, and leaves when it becomes suspicious or stops finding anything new, which is exactly what the rules of Section~\ref{subsubsec:judge} model. A time limit would also bias the comparison against deception, because the sessions it cuts short are those in which deception works best, since an engaged attacker naturally stays longer. Sessions therefore end only when the attacker decides to stop, whether through detection, loss of interest, or exhausting its techniques. The design models a single continuous visit, while a real attacker may also return to the same host over several days; how deception should keep a returning attacker engaged is left as future work.

\subsection{Metrics}
\label{subsec:metrics}

Each session records a set of measurements grouped into four families: engagement, outcome, deception effort, and credibility. The first two describe what the attacker did and why it stopped, the third verifies what the defender actually did, and the fourth tracks how genuine the host looked throughout the session.

\emph{Engagement} captures how much of the attacker's activity the host absorbed, which is the quantity a honeypot exists to maximize. The number of commands issued is the primary measure, since it reflects what the attacker chose to do, whereas session duration is recorded as a secondary measure because it also depends on how fast the models answer on this hardware. The number of unique commands and of unique ATT\&CK techniques tells sustained exploration apart from the repetition of a narrow set of actions, and the number of techniques requested from the reserve of Section~\ref{subsubsec:planner} shows sessions in which the host kept the attacker interested beyond its assigned pool.

\emph{Outcome} records which of the three endings of Section~\ref{subsubsec:planner} closed each session: detection, loss of interest, or exhaustion of the techniques, which is the outcome the defender seeks. Detection and loss of interest are both failures of the defence, yet they call for different remedies, since the first asks for a more convincing host and the second for a more rewarding one, and recording them separately shows which one each condition suffers from. A further flag records whether suspicion arose at some point during a session without ending it, since an attacker may register doubt and continue regardless.

\emph{Deception effort} records what the defender actually did, and serves as the manipulation check without which the condition labels alone would be unverified. The number of distinct capabilities deployed, the total number of deployment steps, and the mean number of capabilities per reaction confirm that each condition behaved as Section~\ref{subsec:conditions} defines it, and in particular that \textsc{Multi-TTP} and \textsc{Chain} deployed the same budget per reaction, so that the two differ in how their artifacts are arranged rather than in how many are planted each time.

\emph{Credibility} differs from the other families because it is judged rather than counted. The credibility scores assigned by the judge of Section~\ref{subsubsec:judge} play two roles. During the session they drive the detection rule, and once the session is over their mean, together with the number of commands issued before the first suspicious response, checks whether the host stayed convincing under each condition, as Section~\ref{sec:results} reports. In practice the judge gives scores close to either $0$ or $1$ rather than spread across the interval, so the mean credibility of a session indicates whether it contained obvious failures rather than grading realism on a fine scale, and it is read accordingly.

\section{Results}
\label{sec:results}

This section reports what the campaign of Section~\ref{sec:setup} measured. Of its $1152$ sessions, $1150$ completed and recorded their results; the other two lost their records to an infrastructure fault, the honeypot failing to write its data file, which is unrelated to how the sessions went, so each combination of condition and skill level rests on $95$ or $96$ sessions. The results are presented in steps. Before any comparison, a manipulation check confirms that each condition deployed what it was meant to, and that \textsc{Multi-TTP} and \textsc{Chain} plant the same budget per reaction. Section~\ref{subsec:engagement} then shows that the more elaborate the deception, the longer the attacker stays, and that coherence matters most against the most skilled attacker. Section~\ref{subsec:abandonment} turns to why the attacker leaves and contains the central result of this work, the retention of the most skilled attackers. Section~\ref{subsec:realism} checks the realism of the host without relying on the credibility judge, Section~\ref{subsec:generalisation} shows that the result does not depend on the models playing each role, Section~\ref{subsec:statistics} confirms it statistically, also within each model pair, and shows that the sessions whose first techniques revealed artifacts that corroborate one another are the ones that last, and Section~\ref{subsec:cost} shows that a coherent reaction needs about as much model time as an unrelated one. Since the distributions are strongly right-skewed, medians and non-parametric tests are used throughout, as Section~\ref{subsec:statistics} explains.

Before comparing how the attacker behaved, it is necessary to verify that each condition actually deployed what Section~\ref{subsec:conditions} says it deploys, since otherwise a difference between conditions could come from a faulty implementation rather than from the property under study. This verification, known as a manipulation check, relies on the deception effort measurements of Section~\ref{subsec:metrics}, summarised in Table~\ref{tab:exposure}. \textsc{Static} deployed nothing, as intended. \textsc{Same-TTP} deployed a median of two distinct capabilities per session, since all of them come from a single Engage activity and many activities offer only a few capabilities, whereas \textsc{Multi-TTP} and \textsc{Chain} deployed fourteen and eighteen. Both of the latter deploy close to four capabilities per reaction, the maximum length of a scenario, so they plant the same amount of deception every time the defensive system reacts. Over a whole session, however, \textsc{Chain} accumulates more deployment steps, $72$ against $50$, because its sessions last longer and give the defensive system more occasions to react. The treatment per reaction is therefore the same, whereas the exposure accumulated over a session is not, so the comparison also has to be made at equal exposure. The same table does so round by round. Per round of the attacker, \textsc{Multi-TTP} and \textsc{Chain} deploy almost the same number of steps, $7.5$ and $7.9$, and \textsc{Multi-TTP} even deploys slightly more distinct capabilities, $2.0$ against $1.8$. At the very first round, before any difference in session length can build up, an \textsc{Advanced} attacker already abandons less often under \textsc{Chain}, $19\%$ against $33\%$. The same table checks that coherence reached the attacker: under \textsc{Chain} a subject is visible in $264$ of its $288$ sessions, against $1$ under \textsc{Multi-TTP}, where names coincide only by chance, and $61\%$ of the file reads that mention a subject name it in two or more paths or commands, so the free text of the artifacts also points to the others. Since a session contains a median of $18$ reactions but shows only $8$ distinct subjects, several scenarios of a session often share a subject, as Section~\ref{subsubsec:engage} anticipates.

\begin{table}[H]
    \centering
    \caption{Deception deployed under each reactive condition (\textsc{Static} deploys nothing). Upper block: per session, as median [interquartile range]. Middle block: per round, where a round is one technique of the attacker, as a mean over all sessions, which compares the conditions at equal exposure. Lower block: evidence that the shared subject reached the attacker.}
    \label{tab:exposure}
    \small
    \setlength{\tabcolsep}{4pt}
    \begin{tabular}{@{}lccc@{}}
        \toprule
        \textbf{Measure} & \textsc{Same-TTP} & \textsc{Multi-TTP} & \textsc{Chain} \\
        \midrule
        \multicolumn{4}{@{}l}{\emph{Per session}} \\
        Distinct capabilities & 2 [2, 4] & 14 [8, 18] & 18 [14, 21] \\
        Deployment steps & 17 [5, 52] & 50 [20, 87] & 72 [35, 133] \\
        Capabilities per reaction & 2 [2, 4] & 3.9 [3.6, 4] & 4 [3.9, 4] \\
        \addlinespace[2pt]
        \multicolumn{4}{@{}l}{\emph{Per round}} \\
        Deployment steps & --- & 7.5 & 7.9 \\
        Distinct capabilities & --- & 2.0 & 1.8 \\
        Deployment steps, \textsc{Advanced} & --- & 4.7 & 5.8 \\
        Abandonment at the first round, \textsc{Advanced} & --- & 33\% & 19\% \\
        \addlinespace[2pt]
        \multicolumn{4}{@{}l}{\emph{Shared subject}} \\
        Sessions in which a subject is visible & --- & 1 of 288 & 264 of 288 \\
        File reads naming the subject in two or more places & --- & --- & 61\% \\
        Reactions per session, median & --- & --- & 18 \\
        Distinct subjects seen per session, median & --- & --- & 8 \\
        \bottomrule
    \end{tabular}
\end{table}

\subsection{Engagement across the deception ladder}
\label{subsec:engagement}

\begin{figure}[H]
    \centering
    \includegraphics[width=\columnwidth]{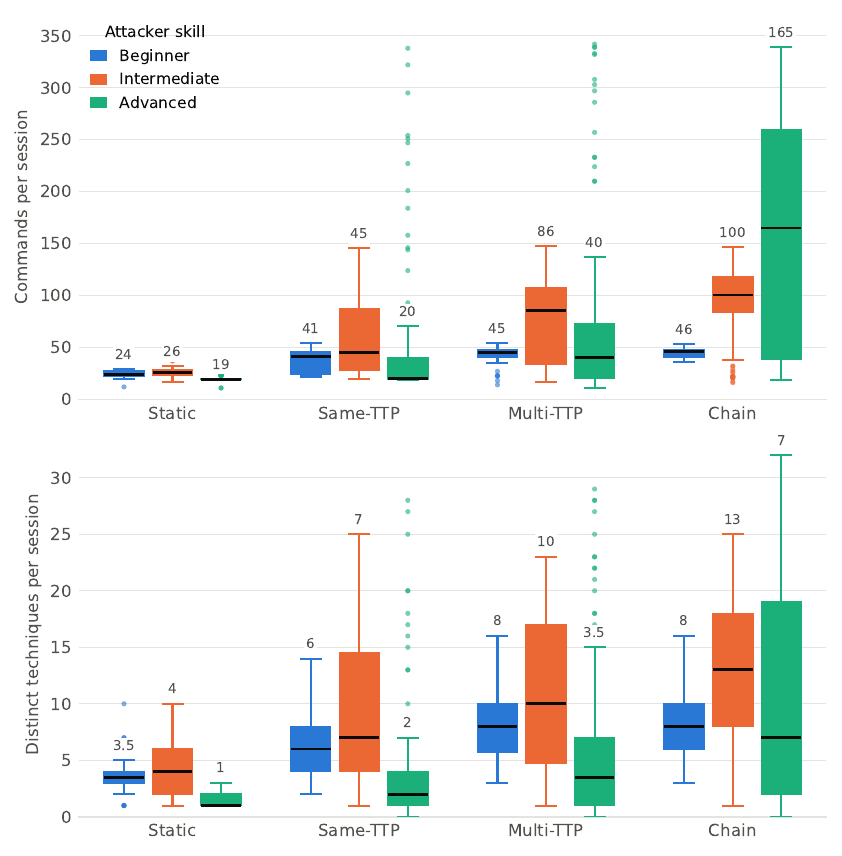}
    \caption{Commands per session (top) and distinct ATT\&CK techniques executed per session (bottom) for each condition and attacker skill level. Each box spans the interquartile range, the line inside it marks the median, which is also written above the box, the whiskers extend to $1.5$ times the interquartile range, and points beyond them are individual sessions.}
    \label{fig:engagement}
\end{figure}

The results show that the more elaborate the deception, the longer the attacker stays. \figurename~\ref{fig:engagement} reports the commands issued per session, which is the primary engagement measure, and the number of distinct ATT\&CK techniques the attacker executed, for each condition and skill level. At every skill level, each condition holds the attacker for more commands than the condition before it, and Section~\ref{subsec:statistics} confirms that this ordering is statistically significant.

\figurename~\ref{fig:engagement} shows the full distribution behind those medians, and each pattern in it follows from the rules of Section~\ref{subsubsec:judge}, which reward each skill level for a different kind of finding. Reading down a column, that is, comparing conditions at one skill level, measures the deception; reading across a row, that is, comparing skill levels under one condition, reflects the attacker model instead. Under \textsc{Static} the boxes of the three levels are narrow and low, between $19$ and $26$ commands, because a host that never reacts offers nothing new, so every attacker leaves after the number of barren rounds that its base patience allows. Against a \textsc{Beginner}, any reaction is enough, since its patience grows with the first rounds that reveal anything at all, so the median rises from $24$ commands under \textsc{Static} to $41$ under \textsc{Same-TTP}, and variety and coherence add little on top of it, with $45$ commands under \textsc{Multi-TTP} and $46$ under \textsc{Chain}. The \textsc{Intermediate} attacker, whose patience grows with the number of different kinds of artifact, gains most from variety, from $45$ to $86$ commands, and a further $14$ commands from coherence. Against an \textsc{Advanced} attacker the picture is the opposite of the \textsc{Beginner}'s. Its patience grows only slightly with variety and mostly with corroboration, so artifacts of a single kind barely change its behavior, from $19$ commands to $20$, variety doubles its median to $40$, and coherence, the only condition that produces corroboration systematically, multiplies it again by four, up to $165$ commands. For a defender, this means that inexpensive deception is enough to keep novice attackers busy, whereas experienced attackers, who are both the most dangerous and the most valuable to observe, are retained only by a scenario that holds together.

Two ingredients of the attacker model explain the rows. The first is the size of the technique pool, six, twelve, or twenty techniques, which sets a ceiling on how many commands a session can reach, so under \textsc{Multi-TTP} and \textsc{Chain} the retained \textsc{Beginner} sessions stop at about $45$ commands because their pool is exhausted, not because they lose interest. The second is what earns patience at each level, described in the previous paragraph. The expert therefore stays longest under \textsc{Chain} not because it receives more patience, since the ceiling is the same at every level, but because it is the attacker with most techniques to try once the host has convinced it to stay.

The number of distinct techniques confirms the same ordering, and shows that longer sessions are not merely the result of an attacker repeating itself. Under \textsc{Static}, an \textsc{Advanced} attacker leaves after a median of a single technique, whereas under \textsc{Chain} it executes seven, so the coherent scenario leads it to try six more ways of exploiting the host, each of which the defender can observe.

\begin{figure}[H]
    \centering
    \includegraphics[width=\columnwidth]{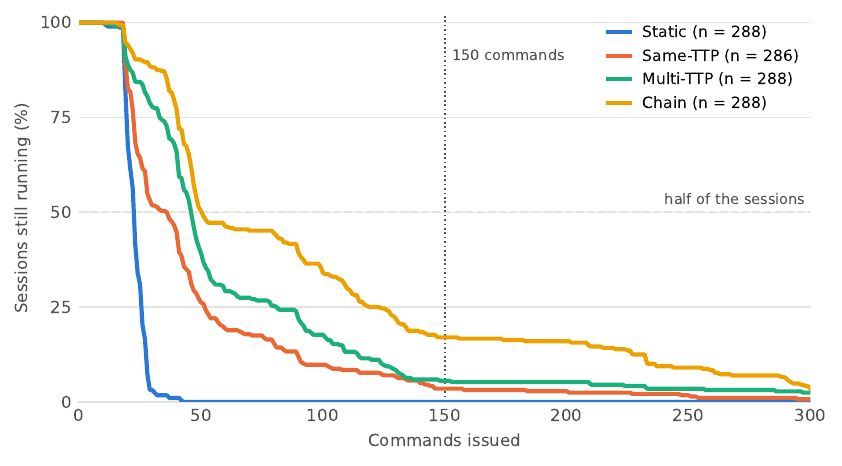}
    \caption{Survival function of commands per session, that is, the percentage of sessions in each condition that went on beyond a given number of commands, computed over all the sessions of that condition ($n$ in the legend), pooling skill levels, models, and persona modes. A session stops when the attacker detects the trap, loses interest, or runs out of techniques. The dotted line marks $150$ commands.}
    \label{fig:survival}
\end{figure}

\figurename~\ref{fig:survival} adds what the medians cannot show, namely when sessions end rather than how long they last on average. Under \textsc{Static}, sessions do not fade out gradually but end almost all at once, since nearly all of them go past $18$ commands while fewer than $4\%$ go beyond $30$. This is what an attacker does when it completes its initial survey, finds nothing that rewards further effort, and leaves, so a host that never reacts only ever observes the opening survey of an intrusion. The reactive conditions keep a share of attackers far beyond that point, and \textsc{Chain} does so most clearly: at $40$ commands, $77\%$ of its sessions are still running, against $66\%$ under \textsc{Multi-TTP} and $45\%$ under \textsc{Same-TTP}, and at $150$ commands it still retains $17\%$, three times as many as \textsc{Multi-TTP}. These long sessions are the valuable ones for a defender, since they are those in which the attacker goes beyond its first survey and reveals how it operates once it believes it has found a worthwhile target.

The differences described so far are large, but sessions within the same condition also vary widely. Section~\ref{subsec:statistics} shows that every step of the ladder is nonetheless statistically significant (Table~\ref{tab:tests}), and that the effect of coherence, from \textsc{Multi-TTP} to \textsc{Chain}, grows with the skill of the attacker: it is not significant against a \textsc{Beginner} and clear against an \textsc{Intermediate} or an \textsc{Advanced} one.

\begin{table}[H]
    \centering
    \caption{Comparison of commands per session between consecutive conditions, using one-sided Mann-Whitney tests. The effect is the rank-biserial correlation: $0$ means that the two conditions are indistinguishable, and $1$ that every session of the more elaborate condition is longer than every session of the other. *** $p < 0.001$; ns, not significant.}
    \label{tab:tests}
    \small
    \begin{tabular}{@{}lrcc@{}}
        \toprule
        \textbf{Comparison} & \textbf{$p$} & & \textbf{Effect} \\
        \midrule
        \multicolumn{4}{@{}l}{Consecutive conditions (all skill levels pooled)} \\
        \textsc{Static} $\rightarrow$ \textsc{Same-TTP}    & $10^{-24}$ & *** & 0.49 \\
        \textsc{Same-TTP} $\rightarrow$ \textsc{Multi-TTP} & $1.2 \times 10^{-7}$ & *** & 0.25 \\
        \textsc{Multi-TTP} $\rightarrow$ \textsc{Chain}    & $1.9 \times 10^{-6}$ & *** & 0.22 \\
        \addlinespace[2pt]
        \multicolumn{4}{@{}l}{The coherence contrast, \textsc{Multi-TTP} $\rightarrow$ \textsc{Chain}, by skill level} \\
        \quad Beginner     & $0.30$ & ns  & 0.04 \\
        \quad Intermediate & $4.7 \times 10^{-4}$ & *** & 0.28 \\
        \quad Advanced     & $7.7 \times 10^{-5}$ & *** & \textbf{0.32} \\
        \bottomrule
    \end{tabular}
\end{table}

That gradient is what the attacker model of Section~\ref{subsubsec:judge} anticipates. A novice is retained by any reactive deception, so no room is left for a scenario to improve on, whereas an experienced attacker discounts unrelated artifacts and responds to evidence that fits together, which is what separates \textsc{Chain} from \textsc{Multi-TTP}.

\begin{figure}[H]
    \centering
    \begin{minipage}[t]{0.48\textwidth}
        \centering
        \includegraphics[width=\textwidth]{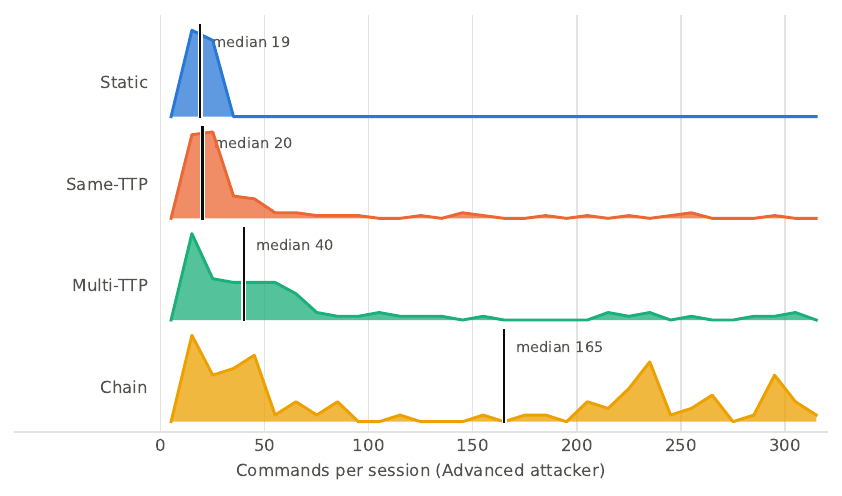}
    \end{minipage}\hfill
    \begin{minipage}[t]{0.48\textwidth}
        \centering
        \includegraphics[width=\textwidth]{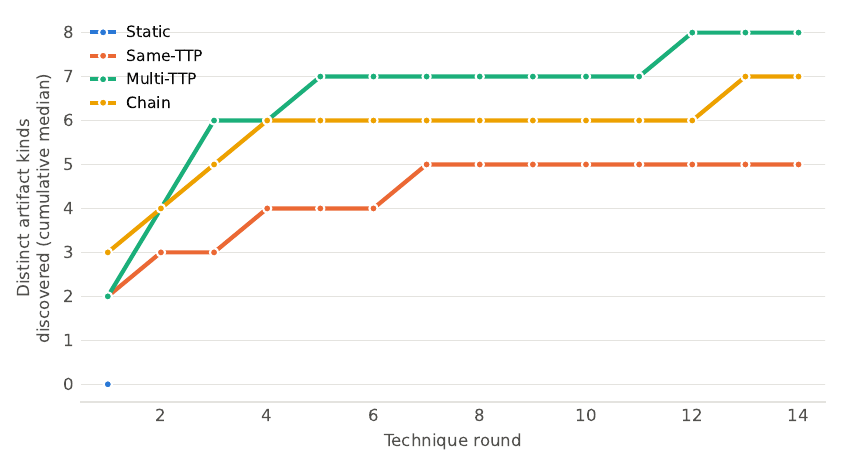}
    \end{minipage}
    \caption{Left: distribution of commands per session against an \textsc{Advanced} attacker, one curve per condition, with the median marked. \textsc{Static} and \textsc{Same-TTP} concentrate almost entirely below $50$ commands, whereas \textsc{Chain} splits into a group of sessions that end early and another that runs beyond $150$ commands. Right: mean number of distinct kinds of artifact discovered after each number of techniques, over all skill levels, counting only the sessions still running at that point. \textsc{Multi-TTP} and \textsc{Chain} reveal new kinds of artifact at practically the same pace.}
    \label{fig:dists_and_discovery}
\end{figure}

The medians and the tests describe a typical session, but against an \textsc{Advanced} attacker the sessions under \textsc{Chain} are far from uniform, as the width of their box in \figurename~\ref{fig:engagement} already suggested. The left panel of \figurename~\ref{fig:dists_and_discovery} shows their distribution in detail, and reveals that \textsc{Chain} does not simply move the sessions towards higher values but splits them into two groups. Of its $96$ sessions against an \textsc{Advanced} attacker, $38$ end before $50$ commands, as under the other conditions, whereas $49$ go beyond $150$. The two groups differ in what the attacker saw before deciding whether to stay. Sessions in the short group ended after a median of one or two techniques, by which point the defender had completed a median of $8$ deployment steps, whereas sessions in the long group went through $220$. An \textsc{Advanced} attacker starts with a base patience of a single round, so when its first rounds do not reveal artifacts that corroborate one another it leaves before the scenario has had time to unfold, whereas once it does find one its patience grows and it stays for a very long time. How quickly the defender delivers that first scenario also depends on the model it runs, since \texttt{llama3.1} defenders account for $21$ of the $49$ long sessions but only $2$ of the $38$ short ones, a dependence that Section~\ref{subsec:generalisation} examines. In practice, therefore, \textsc{Chain} does not retain every expert, but the experts it does retain stay for a very long time, and its first reactions decide which group a session falls into.

The right panel of \figurename~\ref{fig:dists_and_discovery} shows why the advantage of \textsc{Chain} cannot be explained by variety. It gives, after each number of techniques executed, the mean number of different kinds of artifact the attacker has discovered, counting only the sessions still running at that point, so that all conditions are compared after the same exposure. \textsc{Static} stays at zero, since nothing is planted, and \textsc{Same-TTP} stays clearly below the others, since all its artifacts come from a single activity. \textsc{Multi-TTP} and \textsc{Chain}, however, follow almost the same curve: after three techniques the attacker has seen $3.3$ kinds under \textsc{Multi-TTP} and $3.9$ under \textsc{Chain}, and after nine, $6.3$ and $5.9$. Both conditions therefore show the attacker the same variety at the same pace, and \textsc{Chain} sessions accumulate more kinds in total only because they last longer. What separates them is not how many kinds of artifact the attacker finds, but whether its findings refer to one another, and Section~\ref{subsec:statistics} shows that early corroboration is associated with longer sessions and is far more frequent under \textsc{Chain}.

This does not make \textsc{Multi-TTP} a poor option. Its capabilities are not random actions, since they are drawn from the Engage activities that answer the technique the attacker has just used, and they are therefore relevant to what the attacker is doing. \textsc{Multi-TTP} outperforms \textsc{Same-TTP} at every skill level (\figurename~\ref{fig:engagement}), and it is the condition that shows the attacker the greatest variety. Since it needs no curated catalogue of scenarios, it could also serve as a fallback whenever no scenario fits the identified technique or the role of the host. What it cannot do is convince an attacker who checks whether its findings fit together, and against such an attacker the curated scenarios of \textsc{Chain} make the difference.

\subsection{Abandonment behavior by attacker skill}
\label{subsec:abandonment}

The previous subsection measured how long the attacker stayed, and this one turns to why it left, which matters just as much, since the two ways of leaving described in Section~\ref{subsec:metrics} call for different remedies. Table~\ref{tab:abandonment} reports, for each condition and skill level, the percentage of sessions in which the attacker left the host before using up its techniques. Two patterns appear in it. Reading down each column, fewer attackers leave as the deception becomes more elaborate, and reading along each row, more attackers leave as their skill grows, as the design intends, since an experienced attacker has less patience with a host that does not reward it.

\begin{table}[H]
    \centering
    \caption{Percentage of sessions in which the attacker left the host, either by detection or by loss of interest, by condition and attacker skill. Each cell contains $95$ or $96$ sessions, and lower values are better for the defender.}
    \label{tab:abandonment}
    \small
    \begin{tabular}{@{}lcccc@{}}
        \toprule
        & \multicolumn{3}{c}{\textbf{Attacker skill}} & \\
        \cmidrule(lr){2-4}
        \textbf{Condition} & \textbf{Beginner} & \textbf{Intermediate} & \textbf{Advanced} & \textbf{Overall} \\
        \midrule
        \textsc{Static}    & 99\% & 100\% & 100\% & 99.7\% \\
        \textsc{Same-TTP}  & 43\% & 78\%  & 92\%  & 71\% \\
        \textsc{Multi-TTP} & 6\%  & 60\%  & 86\%  & 51\% \\
        \textsc{Chain}     & \textbf{0\%} & \textbf{25\%} & \textbf{52\%} & \textbf{26\%} \\
        \bottomrule
    \end{tabular}
\end{table}

Under \textsc{Static} the attacker left in $287$ of $288$ sessions, which shows that an emulated host on its own, however faithfully it answers, does not hold an attacker's interest, and this is the situation of a conventional honeypot. The last row of the table contains the central result of this work. Against the most demanding attacker, the \textsc{Advanced} one, the coherent scenarios of \textsc{Chain} retain $48\%$ of the sessions, against only $14\%$ under \textsc{Multi-TTP}, which deploys the same budget per reaction without a scenario. Scenario coherence therefore more than triples the number of experts that stay, and it does so against the attackers that are hardest to retain and most valuable to observe. Against \textsc{Intermediate} attackers it retains $75\%$ of the sessions, against $40\%$ under \textsc{Multi-TTP}, and against \textsc{Beginner} ones it retains every session. Coherence does not retain every expert, since about half of them still leave, which is consistent with the two groups of sessions observed in Section~\ref{subsec:engagement}: the experts who leave are mostly those whose first rounds did not yet show them a scenario.

This result rests on an assumption of the attacker model, namely that an experienced attacker values evidence that corroborates itself (Section~\ref{subsubsec:judge}), and corroboration is precisely what \textsc{Chain} is designed to produce. How much the result depends on that assumption can be measured without new sessions. The loss-of-interest rule only decides when a session ends, since neither the planner nor the defensive system reads it before that moment, so any variant of the rule that never grants more patience than the rule as designed ends each session at the same round or earlier, and everything before that round is exactly what was recorded. Replaying the rule as designed on the recorded rounds reproduces the outcome of all $1150$ sessions, and Table~\ref{tab:replay} gives the outcome of its variants. With a rule that ignores corroboration altogether and rewards only variety, at every level, \textsc{Chain} still retains more attackers than \textsc{Multi-TTP} against a \textsc{Beginner} and an \textsc{Intermediate} attacker. Against an \textsc{Advanced} attacker the advantage does depend on corroboration: it disappears when corroboration has no weight, and it is already significant when the attacker gives it a quarter of the weight of the rule as designed. \textsc{Chain} is not worse than \textsc{Multi-TTP} under any variant. The central result therefore holds for any experienced attacker that values corroboration even moderately, and what depends on how much it values it is the size of the advantage, not its direction.

\begin{table}[H]
    \centering
    \caption{Abandonment under variants of the loss-of-interest rule, replayed exactly on the recorded sessions. The weight multiplies the corroboration term of the \textsc{Advanced} branch of Equation~\ref{eq:gamma}, where $1$ is the rule as designed; the variety-only rule replaces every branch with $\min(\kappa_t, 3)$ and ignores corroboration. $p$: one-sided Fisher exact test of \textsc{Chain} against \textsc{Multi-TTP}. Each cell contains $96$ sessions.}
    \label{tab:replay}
    \small
    \begin{tabular}{@{}llccc@{}}
        \toprule
        \textbf{Variant} & \textbf{Attacker} & \textsc{Multi-TTP} & \textsc{Chain} & \textbf{$p$} \\
        \midrule
        Corroboration weight $1$ (as designed) & \textsc{Advanced} & 86\% & \textbf{52\%} & $1.7 \times 10^{-7}$ \\
        Corroboration weight $0.75$ & \textsc{Advanced} & 88\% & \textbf{55\%} & $5.2 \times 10^{-7}$ \\
        Corroboration weight $0.5$ & \textsc{Advanced} & 89\% & \textbf{58\%} & $1.5 \times 10^{-6}$ \\
        Corroboration weight $0.25$ & \textsc{Advanced} & 92\% & \textbf{73\%} & $5.4 \times 10^{-4}$ \\
        Corroboration weight $0.1$ & \textsc{Advanced} & 93\% & 86\% & $0.12$ \\
        Corroboration weight $0$ & \textsc{Advanced} & 96\% & 97\% & $0.78$ \\
        \addlinespace[2pt]
        Variety only & \textsc{Beginner} & 23\% & \textbf{6\%} & $8.7 \times 10^{-4}$ \\
        Variety only & \textsc{Intermediate} & 77\% & \textbf{53\%} & $4.0 \times 10^{-4}$ \\
        \bottomrule
    \end{tabular}
\end{table}

\begin{figure}[H]
    \centering
    \includegraphics[width=\columnwidth]{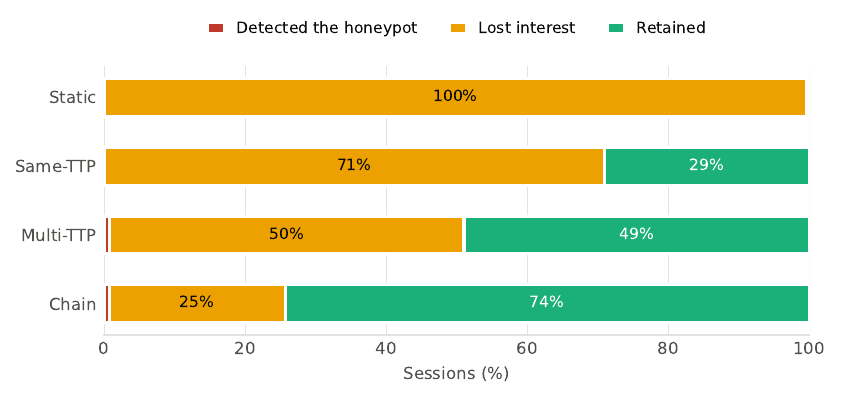}
    \caption{How sessions ended under each condition: detection, loss of interest, or retention, in which the attacker used up all its techniques, including those of the reserve, without leaving. Detection occurred in only four sessions of the whole campaign, two under \textsc{Multi-TTP} and two under \textsc{Chain}, too few to be visible at this scale. The share of retained sessions grows from $0.3\%$ under \textsc{Static} to $74\%$ under \textsc{Chain}, and all of that growth comes from sessions that would otherwise have ended in loss of interest.}
    \label{fig:outcomes}
\end{figure}

\figurename~\ref{fig:outcomes} separates the two ways of leaving. Under the detection rule of Equation~\ref{eq:detection}, detection was practically absent, since no session ended in detection under \textsc{Static} or \textsc{Same-TTP}, and only two did so under each of \textsc{Multi-TTP} and \textsc{Chain}, four out of $1150$ in the whole campaign. Every other departure was a loss of interest, from $99.7\%$ of the sessions under \textsc{Static} to $25\%$ under \textsc{Chain}, and every session that ended in neither way was retained until the attacker ran out of techniques, a share that grows from $0.3\%$ to $74\%$.

Since the detection rule involves several parameters chosen in the design, it is worth asking whether such a low detection rate is merely a consequence of those choices. The rule has no effect on a session until it fires, so it can be replayed on the credibility scores recorded for every session with different parameters, which reproduces exactly the four detections of the campaign and shows what other choices would have produced (Table~\ref{tab:sensitivity}). The threshold $\theta$ and the lower bound of two barely matter, because the judge is almost binary and assigns a score of exactly $1$ to $94\%$ of the responses, so raising $\theta$ from $0.4$ to $0.9$, or letting a single suspicious technique suffice, raises detection to at most $1.7\%$ in any condition. What does matter is which responses count. If the responses to the reconnaissance sweep also counted, detection would rise to $10\%$ of the sessions under \textsc{Same-TTP}, $13\%$ under \textsc{Multi-TTP}, and $24\%$ under \textsc{Chain}, while remaining zero under \textsc{Static}.

\begin{table}[H]
    \centering
    \caption{Sensitivity of detection to the design of the detection rule, obtained by replaying the rule on the recorded credibility scores of every session. The first row is the rule as designed ($\theta = 0.4$, at least two suspicious techniques, sweep not counted). The last block gives the overall abandonment rate, by detection or loss of interest, as designed and when the reconnaissance sweep also counts.}
    \label{tab:sensitivity}
    \small
    \setlength{\tabcolsep}{4pt}
    \begin{tabular}{@{}lcccc@{}}
        \toprule
        \textbf{Rule variant} & \textsc{Static} & \textsc{Same-TTP} & \textsc{Multi-TTP} & \textsc{Chain} \\
        \midrule
        \multicolumn{5}{@{}l}{\emph{Sessions ending in detection}} \\
        As designed                        & 0\%   & 0\%   & 0.7\%  & 0.7\% \\
        Threshold $\theta = 0.9$           & 0\%   & 0\%   & 0.7\%  & 1.0\% \\
        One suspicious technique suffices  & 0.3\% & 1.0\% & 1.7\%  & 1.7\% \\
        Reconnaissance sweep also counts   & 0\%   & 9.8\% & 13.2\% & 24.2\% \\
        \addlinespace[2pt]
        \multicolumn{5}{@{}l}{\emph{Sessions ending in abandonment}} \\
        As designed                        & 99.7\% & 71\% & 51\% & \textbf{26\%} \\
        Reconnaissance sweep also counts   & 99.7\% & 78\% & 59\% & \textbf{46\%} \\
        \bottomrule
    \end{tabular}
\end{table}

The origin of that difference is clear from the commands involved. Almost all the suspicious responses to the sweep come from \texttt{ps aux} and \texttt{netstat}, the commands that list processes and connections, which is exactly where planted processes and connections appear. Under \textsc{Static}, where nothing is planted, the judge found the process list suspicious in $4\%$ of the sweeps, against $18\%$ under \textsc{Same-TTP}, $20\%$ under \textsc{Multi-TTP}, and $27\%$ under \textsc{Chain}, whose longer sessions also run the sweep many more times. Under this judge, the emulated host on its own is rarely found suspicious, whereas the artifacts the defender plants on it, and the fabricated processes above all, are where the deception can give itself away, and the more a condition plants, the more of it the attacker can examine. This is a trade-off rather than a contradiction, since every planted artifact gives the attacker a reason to stay and, when it is imperfect, also a reason to suspect. The ladder survives that trade-off, because even if the sweep counted towards detection, abandonment would still fall from $99.7\%$ under \textsc{Static} to $78\%$ under \textsc{Same-TTP}, $59\%$ under \textsc{Multi-TTP}, and $46\%$ under \textsc{Chain}, and \textsc{Chain} would remain the best condition at every skill level. What \textsc{Chain} gains through interest outweighs what it loses through exposure.

Section~\ref{subsec:realism} checks the realism of the host by means that do not depend on this judge, and Section~\ref{sec:discussion} returns to this trade-off between interest and exposure.

The persona factor, the secondary hypothesis of Section~\ref{subsec:campaign}, produced no significant effect. Over the whole campaign, sessions with a role-specific persona were abandoned in $60\%$ of cases against $64\%$ without one, and their commands per session do not differ significantly ($p = 0.27$). The largest difference appears under \textsc{Chain}, where the persona lowers abandonment from $31\%$ to $21\%$, which suggests that a scenario is more effective when the host has a role it can belong to, although with about $144$ sessions per group this difference is not significant ($p = 0.12$). The result fits the decomposition above, since a persona makes the host more plausible, but plausibility was rarely what limited engagement, whereas what retained the attacker was whether the host kept offering new findings, which a persona on its own does not provide.

\subsection{Realism beyond the credibility judge}
\label{subsec:realism}

The credibility judge that drives detection is a language model, so the low detection rate reported above measures that judge under that rule, not the realism of the host as such. Two checks that do not depend on it were therefore applied to the recorded responses, and Table~\ref{tab:realism} summarizes both. The first compares the output of six common commands with the output of the same commands on a clean Ubuntu~22.04 host, and the second asks a model from a family that took no part in the campaign, \texttt{gemma3:12b}, to score $1200$ recorded responses, $150$ for each condition and kind of command, without ending any session.

\begin{table}[H]
    \centering
    \caption{Realism checks that do not rely on the credibility judge. Upper block: share of the responses to six common commands whose format matches a clean Ubuntu~22.04 host, and two defects introduced by the planted artifacts. Lower block: share of $300$ recorded responses per condition scored below $0.4$ by an independent judge, \texttt{gemma3:12b}, and by the original judge.}
    \label{tab:realism}
    \small
    \setlength{\tabcolsep}{4pt}
    \begin{tabular}{@{}lcccc@{}}
        \toprule
        \textbf{Check} & \textsc{Static} & \textsc{Same-TTP} & \textsc{Multi-TTP} & \textsc{Chain} \\
        \midrule
        \multicolumn{5}{@{}l}{\emph{Format matches a real host}} \\
        \texttt{cat /etc/passwd} & 100\% & 100\% & 100\% & 100\% \\
        \texttt{who} & 100\% & 100\% & 100\% & 100\% \\
        \texttt{ps aux} & 100\% & 99\% & 97\% & 98\% \\
        \texttt{ls -la} & 0\% & 0\% & 0\% & 0\% \\
        \texttt{netstat -tulpn} & 0\% & 0\% & 0\% & 0\% \\
        \texttt{systemctl --failed} & 0\% & 13\% & 42\% & 38\% \\
        \addlinespace[2pt]
        \multicolumn{5}{@{}l}{\emph{Defects of the planted artifacts}} \\
        Sessions where a missing \texttt{netstat} still answers & 4\% & 12\% & 16\% & 17\% \\
        Outputs with Python byte literals & 0\% & 0.3\% & 1.5\% & 2.4\% \\
        \addlinespace[2pt]
        \multicolumn{5}{@{}l}{\emph{Responses judged suspicious}} \\
        Independent judge & 19\% & 21\% & 19\% & \textbf{16.7\%} \\
        Original judge & 1.7\% & 2.0\% & 2.7\% & 3.3\% \\
        \bottomrule
    \end{tabular}
\end{table}

The format checks separate the emulator from the deception. The emulator answers \texttt{cat /etc/passwd}, \texttt{who}, and \texttt{ps aux} in the exact format of a real host, but not \texttt{ls -la}, which it answers like a plain \texttt{ls}, nor \texttt{netstat -tulpn}, whose output lacks its header line, writes the protocol in upper case, and has no column for the owning program, and \texttt{systemctl --failed} omits its header when no unit has failed. These defects are the same in every condition, so they do not affect the comparison between conditions, but an attacker that compared the outputs with a real system would notice them, and the credibility judge scored almost all of these responses as fully credible. The planted artifacts add defects of their own, which grow with the amount deployed: sessions in which the attacker found \texttt{netstat} missing and later received a table of connections from it, and outputs containing the literal form of Python byte strings, such as \texttt{b'...'}, which no real file contains.

The independent judge found many more responses suspicious than the original one, and most of those it flagged among the technique commands were answers such as \texttt{command not found} to binaries or shell constructs that the emulator does not implement, so it mostly points to the coverage of the emulator rather than to the deception. It found no significant difference between conditions ($p = 0.52$ between \textsc{Chain} and \textsc{Multi-TTP}), and \textsc{Chain} was the condition with the fewest suspicious responses, so the planted artifacts did not make the host more suspicious to a second judge. The two judges, however, agreed on very few responses, and neither noticed the format errors of \texttt{ls -la} or \texttt{netstat} that the first check finds, which shows that a language-model judge is not a reliable measure of format realism. The comparison between conditions therefore does not depend on which judge scores the responses, whereas the realism of the emulator itself, namely the coverage of commands and the format of a few common tools, is a limitation of this work that Section~\ref{sec:discussion} discusses.

\subsection{Generalization across models}
\label{subsec:generalisation}

The results so far pool the sixteen model pairs, and this subsection checks whether they depend on which models play each role. The question matters for a specific reason. The numeric parameters of the attacker's rules, such as the threshold $\eta$ or the coefficients of Equation~\ref{eq:gamma}, were set in a pilot campaign in which both the defender and the attacker ran \texttt{llama3.1} (Section~\ref{subsubsec:judge}). The ordering of the conditions could therefore, in principle, be a property of how \texttt{llama3.1} behaves rather than of the deception, and fail to appear with other models.

\begin{figure}[H]
    \centering
    \includegraphics[width=\columnwidth]{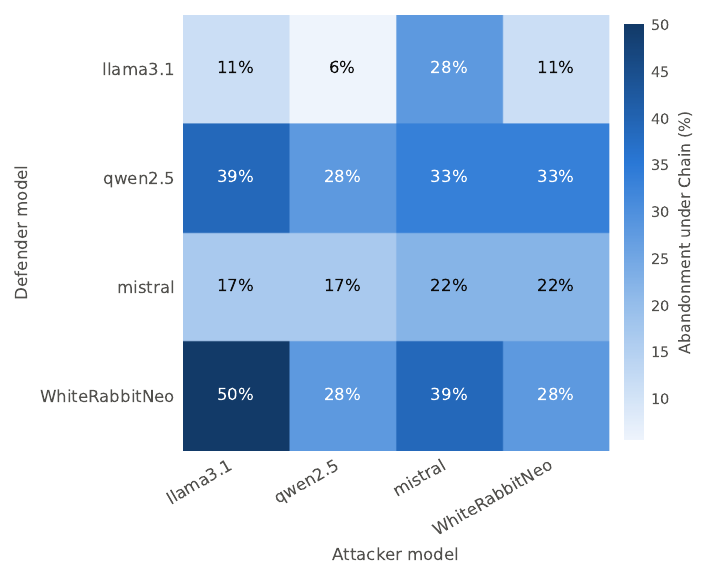}
    \caption{Abandonment rate under \textsc{Chain} for each pair of defender model, in rows, and attacker model, in columns, with $18$ sessions per cell.}
    \label{fig:generalisation_pair}
\end{figure}

The ordering of the conditions holds with every model. Table~\ref{tab:generalisation} shows that, for each of the four attacker models and for each of the four defender models, abandonment decreases at every step of the ladder, from \textsc{Static} to \textsc{Same-TTP}, \textsc{Multi-TTP}, and \textsc{Chain}, including the three models that took no part in the pilot. \figurename~\ref{fig:generalisation_pair} goes down to each of the sixteen pairs and gives their abandonment rate under \textsc{Chain}, which ranges from $6\%$, for a \texttt{llama3.1} defender against a \texttt{qwen2.5} attacker, to $50\%$, for a \texttt{WhiteRabbitNeo} defender against a \texttt{llama3.1} attacker, always far below the $99.7\%$ of \textsc{Static}. Repeating the comparison pair by pair, \textsc{Chain} retains more attackers than \textsc{Multi-TTP} in all sixteen pairs, and none of the other steps of the ladder is reversed in any of them; Section~\ref{subsec:statistics} confirms with tests within each pair that this advantage does not come from how the pairs are mixed.

\begin{table}[H]
    \centering
    \caption{Abandonment rate by condition for each attacker model and for each defender model, pooling all other factors. The ordering holds for every model in both roles, and under \textsc{Chain} the defender's model makes a larger difference than the attacker's.}
    \label{tab:generalisation}
    \small
    \begin{tabular}{@{}lcccc@{}}
        \toprule
        \textbf{Model} & \textsc{Static} & \textsc{Same-TTP} & \textsc{Multi-TTP} & \textsc{Chain} \\
        \midrule
        \multicolumn{5}{@{}l}{\emph{As attacker}} \\
        \texttt{llama3.1}       & 100\% & 71\% & 54\% & 29\% \\
        \texttt{qwen2.5}        & 100\% & 73\% & 44\% & 19\% \\
        \texttt{mistral}        & 100\% & 73\% & 57\% & 31\% \\
        \texttt{WhiteRabbitNeo} & 99\%  & 67\% & 49\% & 24\% \\
        \addlinespace[2pt]
        \multicolumn{5}{@{}l}{\emph{As defender}} \\
        \texttt{llama3.1}       & 100\% & 64\% & 47\% & 14\% \\
        \texttt{qwen2.5}        & 100\% & 72\% & 54\% & 33\% \\
        \texttt{mistral}        & 99\%  & 66\% & 36\% & 19\% \\
        \texttt{WhiteRabbitNeo} & 100\% & 82\% & 67\% & 36\% \\
        \bottomrule
    \end{tabular}
\end{table}

The level of abandonment does vary between models, and it is worth asking which role that variation comes from. Under \textsc{Chain}, abandonment ranges from $14\%$ to $36\%$ depending on the defender's model, and from $19\%$ to $31\%$ depending on the attacker's model. Individual cells of the matrix vary in both directions, since each one contains only $18$ sessions, but when the sessions are pooled, the defender's model has a significant effect on abandonment under \textsc{Chain} (chi-square test, $p = 0.005$), whereas the attacker's model does not ($p = 0.39$), and the same holds over the three reactive conditions together ($p = 9.7 \times 10^{-6}$ against $p = 0.28$). The model that matters is therefore the one that generates the deception, and among those tested, \texttt{llama3.1} retains the most attackers as a defender, which is also why \texttt{llama3.1} defenders accounted for most of the long expert sessions of Section~\ref{subsec:engagement}. One caution applies, since the system was developed with \texttt{llama3.1}, and part of its advantage as a defender may come from that familiarity rather than from the model itself.

\subsection{Statistical analysis}
\label{subsec:statistics}

The previous subsections compared the conditions mainly through medians and percentages. This subsection gathers the statistical tests behind those comparisons and explains what each one adds, so that the reader can judge how firmly the results stand. The tests answer three simple questions: whether the differences between conditions could be due to chance, whether they follow the order of the ladder and hold within each model pair, and whether early corroboration is associated with longer sessions. The size of the campaign helps with all three, since its $1150$ sessions, an order of magnitude more than previous evaluations of generative honeypots (Section~\ref{subsec:campaign}), allow even moderate differences to be told apart from chance.

The tests were chosen to suit the data. The number of commands per session is not spread evenly around its average, since in several cells the standard deviation exceeds the mean, most clearly for \textsc{Same-TTP} against an \textsc{Advanced} attacker, with a mean of $54$ commands and a standard deviation of $73$. This happens because most sessions in such a cell end almost at once while a few run for a long time, so the mean describes no real session. The analysis therefore relies on medians and on rank-based tests, which only ask whether the sessions of one condition tend to be longer than those of another and make no assumption about the shape of the data. Table~\ref{tab:allstats} lists every test, the question it answers, and what its result means.

\begin{table}[H]
    \centering
    \caption{Statistical tests used in this work, with the question each one answers and its reading. All tests use the $1150$ completed sessions, except the rows on early corroboration, which use the reactive conditions, the bootstrap interval, which uses the \textsc{Advanced} attacker, and the tests within model pairs, which compare \textsc{Multi-TTP} and \textsc{Chain}.}
    \label{tab:allstats}
    \small
    \setlength{\tabcolsep}{4pt}
    \begin{tabular}{@{}>{\raggedright\arraybackslash}p{4.2cm}>{\raggedright\arraybackslash}p{2.7cm}>{\raggedright\arraybackslash}p{3.8cm}>{\raggedright\arraybackslash}p{4.0cm}@{}}
        \toprule
        \textbf{Question} & \textbf{Test} & \textbf{Result} & \textbf{Reading} \\
        \midrule
        Do the conditions differ in the commands issued? & Kruskal-Wallis~\cite{kruskal1952} & $H = 386.2$, $p = 2.2 \times 10^{-83}$ & Yes, clearly \\
        Does each condition retain the attacker longer than the previous one? & One-sided Mann-Whitney~\cite{mann1947} & $p \le 1.9 \times 10^{-6}$ at every step (Table~\ref{tab:tests}) & Yes, at every step of the ladder \\
        Do the commands issued follow the order of the ladder? & Spearman correlation~\cite{spearman1904} & $\rho = 0.57$, $p = 3.1 \times 10^{-98}$ & The higher the condition, the more commands \\
        Does abandonment follow the order of the ladder? & Spearman correlation & $\rho = -0.56$, $p = 1.2 \times 10^{-94}$ & The higher the condition, the fewer departures \\
        Does the condition affect whether the attacker leaves? & Chi-square~\cite{pearson1900} with Cram\'er's $V$~\cite{cramer1946} & $\chi^2(3) = 358.3$, $p = 2.4 \times 10^{-77}$, $V = 0.56$ & Yes, with a strong association \\
        Is early corroboration associated with longer sessions? & One-sided Mann-Whitney & $p < 10^{-8}$ in each reactive condition (\figurename~\ref{fig:earlycorr}) & Yes, in every reactive condition \\
        Does it hold where the rule ignores corroboration? & Logistic regression, \textsc{Beginner} and \textsc{Intermediate} & coefficient $-1.01$, $p = 0.007$ & Yes, beyond the amount and variety found \\
        How precise is the abandonment of \textsc{Chain} against an expert? & Bootstrap interval~\cite{efron1979} & $[42\%, 62\%]$, against $[79\%, 93\%]$ for \textsc{Multi-TTP} & The two intervals do not overlap \\
        Does \textsc{Chain} retain more attackers within each model pair? & Cochran-Mantel-Haenszel~\cite{mantel1959} and sign test & CMH $p = 8.2 \times 10^{-10}$; \textsc{Chain} better in $16$ of $16$ pairs, $p = 1.5 \times 10^{-5}$ & Yes, not an effect of how pairs are mixed \\
        Does it hold by skill level within pairs? & Cochran-Mantel-Haenszel & \textsc{Beg.} $p = 0.19$, \textsc{Int.} $p = 1.5 \times 10^{-6}$, \textsc{Adv.} $p = 2.8 \times 10^{-6}$ & Yes for \textsc{Intermediate} and \textsc{Advanced} \\
        \bottomrule
    \end{tabular}
\end{table}

The first two rows show that the four conditions differ in how long they retain the attacker, and that each one retains it longer than the condition before it, including the step that carries the central claim of this work, from \textsc{Multi-TTP} to \textsc{Chain} ($p = 1.9 \times 10^{-6}$, Table~\ref{tab:tests}). Since a $p$-value only says whether a difference exists, Table~\ref{tab:tests} also reports how large it is, through the rank-biserial correlation, which has a direct reading: an effect $r$ means that a session drawn at random from the more elaborate condition is longer than one drawn from the other in a proportion $(1 + r)/2$ of cases, counting ties as half. The largest effect, $0.49$, appears at the first step, where a session of the reactive condition is longer in $75\%$ of cases, as expected at the step where the host starts reacting at all. The effect of coherence over the whole campaign is smaller, $0.22$, or $61\%$ of cases, but it mixes attackers of all levels, and the lower half of Table~\ref{tab:tests} shows that it grows with their skill: there is no detectable difference against a \textsc{Beginner} (effect $0.04$, $p = 0.30$), whereas against an \textsc{Intermediate} attacker a \textsc{Chain} session is longer in $64\%$ of cases (effect $0.28$) and against an \textsc{Advanced} one in $66\%$ (effect $0.32$). These contrasts by skill level, rather than the ordering of the whole campaign, are what the claim of this work rests on. The next two rows add something that those comparisons cannot show on their own. A significant difference would also appear if the conditions were ordered differently, for instance if \textsc{Multi-TTP} retained more attackers than \textsc{Chain}, whereas Spearman's correlation measures whether the results improve steadily with the position of the condition in the ladder, and both the commands issued and the abandonment rate do. The fifth row repeats the comparison for the decision to leave, which is a yes-or-no outcome and is therefore tested with a chi-square test, and Cram\'er's $V$, which measures the strength of the association on a scale from $0$ to $1$, reaches $0.56$, a strong value.

\begin{figure}[H]
    \centering
    \includegraphics[width=\columnwidth]{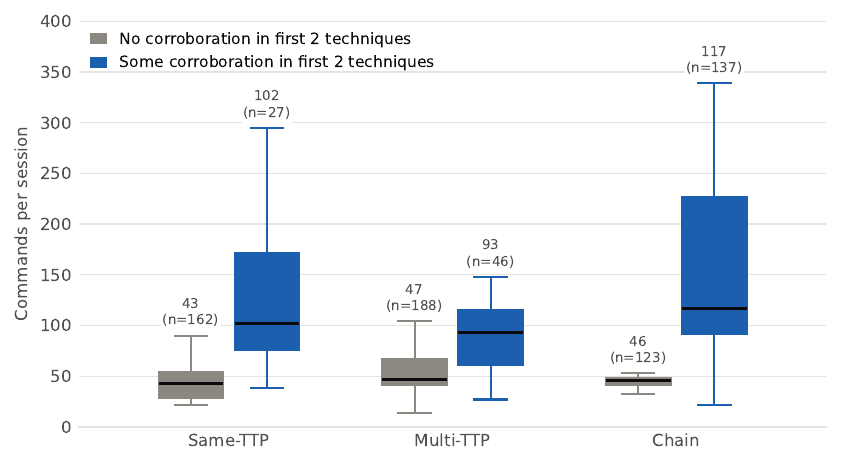}
    \caption{Commands per session under the reactive conditions, separating the sessions in which the first two techniques already produced some corroboration from those in which they did not, among the sessions that executed more than two techniques. The number above each box is its median, and $n$ the number of sessions. In every condition early corroboration more than doubles the length of the session, and \textsc{Chain} produces it in more than half of its sessions.}
    \label{fig:earlycorr}
\end{figure}

The next two rows look at why \textsc{Chain} might retain attackers for longer. The comparisons above treat each condition as a whole, which leaves open the possibility that \textsc{Chain} differs from \textsc{Multi-TTP} in some other, unmeasured way that happens to go together with coherence. A more direct test looks at the corroboration that each attacker actually found, that is, at whether the artifacts it discovered shared their names with one another. Comparing the total corroboration of each session with its length would be misleading, since a longer session finds more corroboration merely by lasting longer. The test therefore looks only at the first two techniques of each session, and asks whether the sessions in which those techniques already produced some corroboration went on to last longer than those in which they did not (\figurename~\ref{fig:earlycorr}). They did, in every reactive condition and by a wide margin, since the median session grows from $43$ to $102$ commands under \textsc{Same-TTP}, from $47$ to $93$ under \textsc{Multi-TTP}, and from $46$ to $117$ under \textsc{Chain}. The effect appears even under \textsc{Same-TTP} and \textsc{Multi-TTP}, where corroboration arises only when two unrelated artifacts happen to share a name, so the association does not come from some other property of \textsc{Chain}. It is an association rather than a proof of cause, and part of it could come from the rule itself, since the \textsc{Advanced} branch of Equation~\ref{eq:gamma} rewards corroboration directly. It also holds, however, where the rule does not read corroboration at all: for \textsc{Beginner} and \textsc{Intermediate} attackers, a logistic regression of abandonment on early corroboration, adjusted for the amount and the variety of what the first two techniques revealed and for the condition, gives early corroboration a coefficient of $-1.01$ ($p = 0.007$), so it lowers the odds of leaving even when patience does not depend on it. What \textsc{Chain} changes is how often early corroboration happens, since its scenarios produce it in $53\%$ of the sessions, against $20\%$ under \textsc{Multi-TTP} and $14\%$ under \textsc{Same-TTP}, which is consistent with the mechanism proposed in Section~\ref{sec:design}.

The last row shows how precise the central figure of Section~\ref{subsec:abandonment} is. The abandonment rate of \textsc{Chain} against an \textsc{Advanced} attacker, $52\%$, has a $95\%$ bootstrap interval of $[42\%, 62\%]$ over $10{,}000$ resamples, against $[79\%, 93\%]$ for \textsc{Multi-TTP} in the same cell. The two intervals do not overlap, which supports the claim that coherence reduces the abandonment of experienced attackers, although they are wide enough to show that $52\%$ should not be read as a precise value.

The tests above pool the sixteen model pairs, which could hide a result that holds only for some of them or that comes from how the pairs are mixed. The last two rows of Table~\ref{tab:allstats} therefore compare \textsc{Chain} with \textsc{Multi-TTP} within each pair. The Cochran-Mantel-Haenszel test combines the sixteen within-pair comparisons of abandonment into one, and a sign test counts the pairs in which \textsc{Chain} retains more attackers, which it does in all sixteen. Both confirm the advantage of \textsc{Chain} against \textsc{Intermediate} and \textsc{Advanced} attackers, while against a \textsc{Beginner} it remains not significant, as in the pooled analysis.

Two contrasts are not significant, and both mark the limits of the claim rather than contradict it. Against a \textsc{Beginner}, \textsc{Chain} and \textsc{Multi-TTP} retain the attacker equally ($p = 0.30$, Table~\ref{tab:tests}): any reactive deception already keeps a novice engaged, so a scenario has nothing left to add. The persona has no significant effect on engagement ($p = 0.27$, Section~\ref{subsec:abandonment}): its purpose is to prevent inconsistencies that would give the host away, and since detection ended only four sessions (Section~\ref{subsec:abandonment}), plausibility was rarely what limited engagement in this campaign. Finally, no correction for multiple comparisons is applied to the main contrasts, since they were fixed in advance by the design of Section~\ref{subsec:conditions} rather than chosen after seeing the data, and their $p$-values are in any case several orders of magnitude below any threshold that such a correction would impose.

\subsection{Computational cost}
\label{subsec:cost}

More elaborate deception requires more work from the language models, and this subsection measures it as the duration and number of model calls of each condition, which matters both for deployment, to know what each engaged attacker costs, and for the experiment, whose comparison must not be distorted by conditions that compute more than others.

The measurements come from a separate campaign of $96$ sessions in which every model call was timed. To keep that campaign short, it covered two model pairs instead of sixteen, chosen to represent the two situations of the main campaign: \texttt{llama3.1} against itself, the pair with which the system was calibrated and in which both sides run the same model, and \texttt{mistral} defending against \texttt{qwen2.5}, two models from different developers that took no part in the calibration. Both pairs ran every condition, skill level, and persona mode. Absolute times depend on the hardware described in Section~\ref{subsec:models} and on how the models are served, which can change inference latency by an order of magnitude~\cite{mlperf2025}, so the values below are best read as the relative cost of each component and condition.

\figurename~\ref{fig:cascade} places the three time scales that describe the system on a single logarithmic axis, since they differ by seven orders of magnitude:

\begin{figure}[H]
    \centering
    \includegraphics[width=\columnwidth]{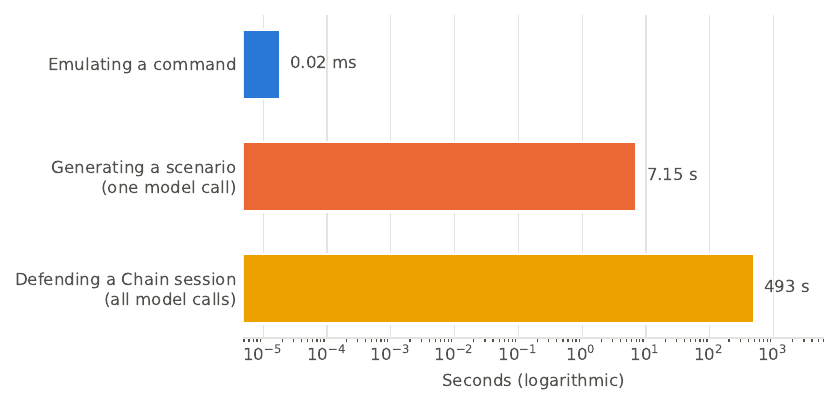}
    \caption{The three time scales of the system, on a logarithmic axis. Emulating a command is the median over $8{,}631$ commands of the campaign replayed on fresh emulated hosts; generating a scenario is the median duration of one model call that writes all the artifacts of a scenario; defending a \textsc{Chain} session is the median, over the instrumented \textsc{Chain} sessions, of all the defender's model calls added up.}
    \label{fig:cascade}
\end{figure}

\begin{itemize}
    \item \emph{Emulating a command} takes a median of $0.02$ milliseconds. Turning a command such as \texttt{ls /var/log} into its output involves no language model, since the host is emulated by ordinary code, so the host answers every command at least as fast as a real server would, whatever the condition and whatever the defender is doing. Seen from the attacker, the response to a typical command reaches its client over SSH in less than $0.1$ seconds, the resolution with which the client checks for incoming data, in every condition of the main campaign.
    \item \emph{Generating a scenario} takes a median of $7.15$ seconds, the time of one model call that writes all the artifacts of a scenario. This work is done by the defender alongside the session, separately from the emulation that answers each command.
    \item \emph{Defending a whole \textsc{Chain} session} adds up to a median of $493$ seconds, about eight minutes of model computation, which is what a deployment must provide for each attacker it engages with the complete system.
\end{itemize}
For a real deployment, these values mean that answering commands is never the bottleneck, that a single workstation such as the one of Section~\ref{subsec:models} produces a complete scenario in a few seconds, and that engaging an attacker for a whole session costs minutes of computation on that same workstation, without specialized infrastructure.

Table~\ref{tab:cost} breaks the computation down by component and condition, which shows where the time goes and how each condition spends it. The technique correlator runs on the commands of every condition, so its number of calls simply follows the length of the sessions, from $7.5$ calls per session under \textsc{Static}, whose sessions are the shortest, to between $29.5$ and $38.0$ under the reactive conditions. Under \textsc{Static} it is the only defensive component that runs, since nothing is deployed, which is why that condition costs only $10$ seconds of model time per session. The reactive conditions differ in how they generate their artifacts. \textsc{Same-TTP} and \textsc{Multi-TTP} write the artifacts of each capability with a separate call of $5.38$ seconds, and need $76.5$ and $107.4$ such calls per session. \textsc{Chain}, instead, writes all the artifacts of a scenario in a single call of $7.15$ seconds, of which it makes $30.3$ per session, preceded by the call of the Engage planner that chooses the scenario, and it generates artifacts one capability at a time only in the $15.6$ reactions per session whose scenario contains a single capability.

\begin{table}[H]
    \centering
    \caption{Model calls per session by component and condition. The duration of each call is the median [interquartile range] over all instrumented sessions. Calls per session are means for each condition, and for the attacking system over all sessions. Reactions per session, model time per reaction (planner and artifact generation divided by the number of reactions), and model time per session (all model calls added up) are medians over the sessions of each condition, and for the attacking system the model time is its mean over all sessions.}
    \label{tab:cost}
    \small
    \setlength{\tabcolsep}{3pt}
    \begin{tabular}{@{}lccccc@{}}
        \toprule
        & & \multicolumn{4}{c}{\textbf{Calls per session}} \\
        \cmidrule(l){3-6}
        \textbf{Component} & \textbf{Duration (s)} & \textsc{Static} & \textsc{Same} & \textsc{Multi} & \textsc{Chain} \\
        \midrule
        \multicolumn{6}{@{}l}{\emph{Defensive system}} \\
        Technique correlator                & 2.15 [1.94, 3.08] & 7.5 & 29.5 & 38.0 & 33.2 \\
        Engage planner                      & 2.60 [2.22, 4.97] & 0   & 0    & 0    & 27.3 \\
        Artifacts, whole scenario           & 7.15 [5.42, 8.92] & 0   & 0    & 0    & 30.3 \\
        Artifacts, one capability           & 5.38 [2.76, 8.26] & 0   & 76.5 & 107.4 & 15.6 \\
        \addlinespace[2pt]
        Reactions per session               &                   & 0   & 24.5 & 28   & 28 \\
        Model time per reaction (s)         &                   & --- & 15.5 & 15.6 & 16.2 \\
        Model time per session (s)          &                   & 10  & 283  & 719  & 493 \\
        \midrule
        \multicolumn{6}{@{}l}{\emph{Attacking system, over all sessions}} \\
        Attack planner                      & 1.57 [1.40, 1.97] & \multicolumn{4}{c}{9.4} \\
        Credibility judge, model calls      & 1.44 [1.40, 1.51] & \multicolumn{4}{c}{69.8} \\
        Credibility judge, without a model  & ---               & \multicolumn{4}{c}{53.4} \\
        Follow-up commands                  & 1.63 [1.27, 2.44] & \multicolumn{4}{c}{0.1} \\
        \addlinespace[2pt]
        Model time per session (s)          &                   & \multicolumn{4}{c}{127} \\
        \bottomrule
    \end{tabular}
\end{table}

These two ways of generating artifacts end up costing the same. Per reaction, the three reactive conditions need almost the same model time, a median of $15.5$ seconds under \textsc{Same-TTP}, $15.6$ under \textsc{Multi-TTP}, and $16.2$ under \textsc{Chain} (Table~\ref{tab:cost}), since \textsc{Chain} replaces several short calls, one per capability, with one longer call for the whole scenario plus the call of the planner that chooses it. Over a whole session, \textsc{Chain} shows a lower median than \textsc{Multi-TTP}, $493$ seconds against $719$, but with $24$ sessions per condition in this campaign the difference is not significant (one-sided Mann-Whitney test, $p = 0.21$), and it reflects how many reactions each session happened to contain rather than a cheaper reaction. The justified conclusion is therefore that, in this timing campaign, a coherent reaction needed about as much model time as an unrelated one, while retaining many more attackers (Section~\ref{subsec:abandonment}). The catalogue of scenario templates has a cost of its own, paid once rather than per session, since its fifty-seven templates were written by hand. \textsc{Same-TTP} has the lowest model time per session, $283$ seconds, because its sessions are shorter and contain fewer reactions.

The attacking system, in the lower half of Table~\ref{tab:cost}, needs $127$ seconds of model time per session, less than the defender under any reactive condition, since it only decides and judges, whereas the defender writes new content. Its most frequent component is the credibility judge, which evaluates every response, but only $69.8$ of its evaluations per session reach a model. The other $53.4$ concern commands that produce no output at all, such as a successful \texttt{cd} or \texttt{touch}, which are scored as credible directly, since an empty answer is exactly what a real shell returns for them and there is nothing in it for a model to judge. All the models of the attacking system receive only the text of each command and its output, so the evaluation concerns what an attacker reads on the screen, not the timing of the exchange, which Section~\ref{sec:discussion} discusses as a limitation.

The extra computation of the richer conditions could, in principle, distort the comparison between them, since a session that involves more computation also takes longer in real time, and a longer session might reflect a slower system rather than a more engaging one. Three observations rule this out. First, the primary measure counts commands rather than seconds (Section~\ref{subsec:campaign}), and an \textsc{Advanced} attacker issuing $165$ commands under \textsc{Chain} against $40$ under \textsc{Multi-TTP} (\figurename~\ref{fig:engagement}) performs four times as many actions, however long each one takes. Second, the number of distinct techniques behaves in the same way, rising from $3.5$ to $7$ for an \textsc{Advanced} attacker in the same figure. Third, the decision to leave does not depend on time either, since the loss-of-interest rule of Section~\ref{subsubsec:judge} counts rounds without new findings, so a slow host with nothing to offer is abandoned after exactly as many rounds as a fast one. For these reasons, every claim of this work rests on counts of actions rather than on elapsed time.


\section{Discussion}
\label{sec:discussion}

The campaign of Section~\ref{sec:results} supports a claim narrower than the proposition that deception works, and more useful. Reactive deception of any kind beats a static environment by a wide margin, which the step from \textsc{Static} to \textsc{Same-TTP} establishes at an effect size of $0.49$. Beyond that floor, however, what separates good deception from adequate deception is whether the fabricated evidence coheres, and the comparison carrying that claim keeps the same budget per reaction: \textsc{Multi-TTP} and \textsc{Chain} plant close to four artifacts every time they react and almost the same amount per round, and they differ in whether those artifacts describe one event. Against an \textsc{Advanced} attacker the median session nonetheless grows from $40$ commands to $165$ and abandonment falls from $86\%$ to $52\%$. The data also point to the mechanism, since early corroboration is associated with longer sessions in every reactive condition and is far more frequent under \textsc{Chain} (Section~\ref{subsec:statistics}). Coherence is therefore not a refinement of quantity but a separate axis along which deception can improve, and one that, in the timing campaign, needed about as much model time per reaction as an unrelated deployment (Section~\ref{subsec:cost}). This conclusion holds under an attacker model whose experts value evidence that corroborates itself, but it needs little of that assumption: Table~\ref{tab:replay} shows that against an \textsc{Advanced} attacker the advantage of \textsc{Chain} is significant once corroboration carries a quarter of its designed weight, and that against \textsc{Beginner} and \textsc{Intermediate} attackers \textsc{Chain} remains the best condition even when corroboration is ignored.

What makes that result more than a curiosity is the direction in which it strengthens. Effect sizes in Table~\ref{tab:tests} rise from $0.04$ against a \textsc{Beginner}, where the contrast is not significant, through $0.28$ against an \textsc{Intermediate}, to $0.32$ against an \textsc{Advanced} attacker. Such a gradient is the opposite of what a defence would produce if it merely added noise, since noise fools the inattentive rather than the careful. A novice is retained by any reactive deception at all, which leaves no room for a scenario to improve on, whereas a capable attacker discounts unrelated artifacts and responds specifically to evidence that fits together. For a practitioner the implication is the opposite of what intuition suggests, because coherence matters least against the attackers a honeypot already handles and most against those it would otherwise lose.

A second observation concerns where defensive effort should go. Under the detection rule as designed, four sessions out of $1150$ ended because the attacker concluded it faced a trap, and with nothing planted the emulated host was almost never judged suspicious, so essentially every departure stemmed from the attacker judging the target unrewarding rather than artificial. Under that judge, improving the fidelity of the emulator would therefore address a failure that rarely ended a session, which matters because raising that fidelity is the stated objective of recent generative honeypots~\cite{sladic2024llm,otal2024llmhoneypot}, although Section~\ref{subsec:realism} shows that the emulator still departs from a real host in ways that an attentive attacker could notice. The sensitivity analysis of Section~\ref{subsec:abandonment} nonetheless shows that realism has not stopped mattering, but has moved: when the routine reconnaissance sweep also counts towards detection, detection grows with the amount planted, almost always because of fabricated processes that look out of place in \texttt{ps aux} (Table~\ref{tab:sensitivity}). Deception thus faces a trade-off that a static honeypot never meets: every artifact it plants is both a reason for the attacker to stay and, if imperfect, a clue that the host is a trap, and the more it plants, the more clues it exposes. In this campaign the first effect clearly outweighed the second, since \textsc{Chain} remains the best condition at every skill level even when the sweep counts, but the second sets the direction for improving it. The quantity worth optimising is therefore whether the target keeps offering the attacker reasons to stay, and the realism worth investing in is no longer that of the emulated host but that of what the defender plants on it, starting with the processes, whose generation could be constrained to the formats and names that a genuine process table would show.

Generalisation across models adds a further practical lesson. The ordering of the conditions holds for each of the four models in both roles, including the three that took no part in the calibration, and \textsc{Chain} retains more attackers than \textsc{Multi-TTP} in all sixteen model pairs. The models do differ in how much they retain, and that difference comes from the defender's side, since the defender's model has a significant effect on abandonment under \textsc{Chain} whereas the attacker's model does not (Section~\ref{subsec:generalisation}). Within the four models tested, a defender allocating limited capacity should therefore invest in the model that fabricates the artifacts rather than try to anticipate the sophistication of the attacker it might face.

The system itself has limitations that suggest how to improve it. The Technique Correlator classifies each command with a short window of history, so a wrong classification carries over to the following steps, intent spread over a long stretch of the session is only partly visible, and the first commands have little context; keeping the few most likely techniques with their confidence, revising the choice when later commands contradict it, and passing a short summary of the techniques already identified would address all three while keeping prompts short. Commands below the confidence threshold, such as a previously unseen exploit, and techniques with no counterpart in MITRE Engage receive no deception at all; a low-cost reaction that does not depend on the technique, drawn from activities such as \emph{Pocket Litter} or \emph{Burn in}, or the activities of another technique of the same tactic, would keep the host evolving in those cases. The Artifact Factory guarantees the form of each artifact but not that its free content fits the persona, which giving the generating prompt the installed packages and services, and rejecting artifacts that name software the host does not run, would fix. Finally, the emulator answers a few common tools, such as \texttt{ls -la} and \texttt{netstat}, in a format that differs from a real Ubuntu host, and does not implement some binaries and shell constructs, so correcting those formats and extending the coverage of commands is a direct improvement that no change to the deception would replace.

Two limitations of the evaluation bound how far these conclusions travel. The first concerns who the attacker is, since every attacker in this campaign is a language model rather than a person, and such an agent neither tires nor improvises beyond its prompt, so its judgement of what is interesting can only stand in for human judgement, not reproduce it. The results establish that coherent deception retains LLM-driven attackers and extend to human operators only under the assumption that both respond to narrative coherence in similar ways, which a study with human red-team participants would settle directly. A related bound concerns what that attacker was equipped to notice, since the credibility judge evaluates the content of responses rather than their timing, and timing is an established fingerprinting vector against honeypots~\cite{defibaugh2004,sok2025honeypots}. An attacker instrumented to measure inter-arrival times might therefore separate the conditions where the present one did not. The second concerns the ceiling the design itself imposes, because under \textsc{Chain} every \textsc{Beginner} session and $75\%$ of \textsc{Intermediate} sessions exhausted their assigned pool of techniques rather than losing interest, so for those sessions the measured engagement is only a lower bound, since they might have continued had more techniques been available. This limit affects mainly the cells in which deception performed best, so the advantage of the richer conditions is, if anything, larger than reported, and larger technique pools would reveal it.

None of these limitations favours one condition over another in the central comparison, since \textsc{Multi-TTP} and \textsc{Chain} share all of them and differ in how their artifacts are arranged.

\section{Conclusions}
\label{sec:conclusions}
This work investigated whether narrative coherence improves the effectiveness of cyber deception against LLM-driven attackers. To this end, we presented a honeypot framework that generates scenarios composed of mutually referential, MITRE Engage-based artifacts in response to observed attacker behaviour. The experimental campaign, comprising 1,150 completed sessions across four deception conditions, four attacker models, four defender models, and three skill levels, showed a consistent improvement in engagement as deception became more structured. In particular, \textsc{Chain} outperformed \textsc{Multi-TTP} for Intermediate and Advanced attackers, whereas the difference was not statistically significant for Beginners. For Advanced attackers, the median number of commands increased from 40 to 165, while abandonment decreased from 86\% to 52\%. These results support the value of organizing deceptive artifacts into coherent scenarios instead of generating them independently within the evaluated environment and against the attacker model.

Moreover, the results highlight that deception effectiveness depends on the apparent realism of the host and also on its ability to provide meaningful findings that sustain further adversarial exploration. Under the configured detection rule, most sessions ended because the attacker lost interest rather than the detection of the honeypot itself. Additionally, the advantage of \textsc{Chain} over \textsc{Multi-TTP} persisted across all sixteen model pairs, while the timing evaluation found comparable inference time per reaction for both conditions. Nevertheless, the results should be interpreted within the scope of the experimental design, e.g.., the attackers were LLM agents or that the emulated host had limited command coverage.

Future work will focus on improving technique classification, ensuring consistency between generated artifacts and the emulated host, and extending command coverage. The evaluation will also be expanded to include human red-team participants, larger models, timing-aware attackers, and larger technique pools. Finally, incorporating attacker motivation, repeated visits, coordinated attackers, and concurrent sessions across multiple hosts will help assess the applicability of narrative deception in more complex environments.

\section*{Data and code availability}

The code of the defensive and attacking systems, the catalogue of scenario templates and the mapping of Engage activities to capabilities, every prompt, and the configuration of the campaign of Section~\ref{subsec:campaign}, from which the seed of every session is derived, are publicly available~\cite{llmhoney_engage}. The records of all sessions of the campaign and the scripts that produce the tables and figures of this paper are available from the corresponding author on reasonable request.

\section{Acknowledgments} 

This work was supported by Seneca Foundation in Region of Murcia (Spain) (22943\\/FPI/25), co-funded by Scorpion Cybertechnologies.

\bibliographystyle{elsarticle-num} 
 \bibliography{cas-refs}





\end{document}